\documentclass[aps,prl,reprint,superscriptaddress,longbibliography]{revtex4-1}

\newcommand{\bra}[1]{\langle #1 |}
\newcommand{\ket}[1]{| #1 \rangle }

\usepackage{amsmath}
\usepackage{amstext}
\usepackage{amssymb}
\usepackage{graphicx}
\usepackage{import}
\usepackage{appendix}
\usepackage{placeins}

\usepackage[pdfpagemode=UseNone,pdfstartview=FitH,colorlinks=true,linkcolor=blue,citecolor=blue,urlcolor=blue]{hyperref}
\usepackage[all]{hypcap}
\usepackage{subfigure}

\begin{document}

\title{Anneal-path correction in flux qubits}

\newcommand{\ngc}{Northrop Grumman Corporation, Linthicum, Maryland 21090, USA}
\newcommand{\uscee}{Department of Electrical Engineering, University of Southern California, Los Angeles, California 90089, USA}
\newcommand{\uscC}{Center for Quantum Information Science \& Technology, University of Southern California, Los Angeles, California 90089, USA}
\newcommand{\uscphys}{Department of Physics, University of Southern California, Los Angeles, California 90089, USA}
\newcommand{\uscchem}{Department of Chemistry, University of Southern California, Los Angeles, California 90089, USA}

\author{Mostafa Khezri} \thanks{mkhezri@usc.edu}
\affiliation{\uscee}
\affiliation{\uscC}
\author{Jeffrey A. Grover}
\affiliation{\ngc}
\author{James I. Basham}
\affiliation{\ngc}
\author{Steven M. Disseler}
\affiliation{\ngc}
\author{Huo Chen}
\affiliation{\uscee}
\affiliation{\uscC}
\author{Sergey Novikov}
\affiliation{\ngc}
\author{Kenneth M. Zick}
\affiliation{\ngc}
\author{Daniel A. Lidar} 
\affiliation{\uscee}
\affiliation{\uscC}
\affiliation{\uscchem}
\affiliation{\uscphys}

\begin{abstract}
Quantum annealers require accurate control and optimized operation schemes to reduce noise levels, in order to eventually demonstrate a computational advantage over classical algorithms.
We study a high coherence four-junction capacitively shunted flux qubit (CSFQ), using dispersive measurements to extract system parameters and model the device.
Josephson junction asymmetry inherent to the device causes a deleterious nonlinear cross-talk when annealing the qubit.
We implement a nonlinear annealing path to correct the asymmetry in-situ, resulting in a substantial increase in the probability of the qubit being in the correct state given an applied flux bias.
We also confirm the multi-level structure of our CSFQ circuit model by annealing it through small spectral gaps and observing quantum signatures of energy level crossings.
Our results demonstrate an anneal-path correction scheme designed and implemented to improve control accuracy for high-coherence and high-control quantum annealers, which leads to an enhancement of success probability in annealing protocols.
\end{abstract}

\maketitle

\section{Introduction}
Quantum annealing (QA) began as a quantum-inspired classical optimization method~\cite{Apolloni1988,Finnila1994,Kadowaki1998} and motivated proposals for adiabatic quantum computing~\cite{Farhi2000,Albash2018,Hauke2020}, an analog model of universal quantum computation~\cite{Aharonov2008}.
Flux qubits~\cite{Mooij1999} are a natural choice for implementing QA, since they exhibit a tiltable double-well potential.
The quantum states are characterized by persistent supercurrents flowing in opposite directions that correspond to the states in each well, and these currents can be mapped onto the binary spin variables used in QA~\cite{Kaminsky2004}.
The qubits are initialized in a potential with a low barrier (i.e., large tunneling) between the two wells and no net persistent current.
Toward the end of the anneal, the potential barrier is raised to reduce the tunneling between the wells, giving qubits a net persistent current.
A measurement of the persistent current direction is made to determine the final qubit state.

The coherence of a flux qubit is affected by a variety of noise sources, in particular flux noise that couples to the qubit via its persistent current $I_\text{p}$.
This can limit the energy relaxation time and coherence time, which for slow flux noise scales roughly as $1/I_\text{p}^2$ and $1/I_\text{p}$, respectively~\cite{Quintana2017,Weber2017}.
D-Wave Systems has performed much of the pioneering work in this field~\cite{Berkley2010,Harris2010,Johnson2010} using niobium-based qubits with relatively high persistent currents ($I_\text{p}\sim 3\mu$A), which limits the relaxation and coherence times to $\sim$ 20 ns~\cite{Ozfidan2020}.
Our work is performed using capacitively-shunted flux qubits (CSFQs)~\cite{You2007,Yan2016} fabricated at MIT Lincoln Laboratory by patterning high-quality aluminum on a silicon substrate.
They are designed to have small persistent currents ($I_\text{p}\sim 170$\,nA) and exhibit $\gtrsim 100$ times longer $T_1$ and $T_2$~\cite{Yan2016,Weber2017,Novikov2018}.

A key challenge for flux qubits is their sensitivity to fabrication variations of the Josephson junction critical currents.
In particular, junctions in a SQUID loop exhibit different critical currents despite identical design. 
This junction asymmetry causes nonlinear crosstalk~\cite{Yoshihara2006} between the qubit control fluxes that, if left uncompensated, has significant adverse effects on operational fidelity.
One mitigation technique is to use compound junctions~\cite{Harris2010}, replacing each junction with a SQUID loop of two junctions.
Flux biasing these loops allows tuning of the effective junctions to achieve nearly identical critical currents.
The trade-off is increased flux noise sensitivity (thus reducing $T_1$ and $T_2$), control overhead for the additional bias lines, and a lengthier crosstalk calibration procedure (which scales quadratically with the number of bias lines).

\begin{figure}[t]
	\begin{center}
		\includegraphics[width=1\columnwidth]{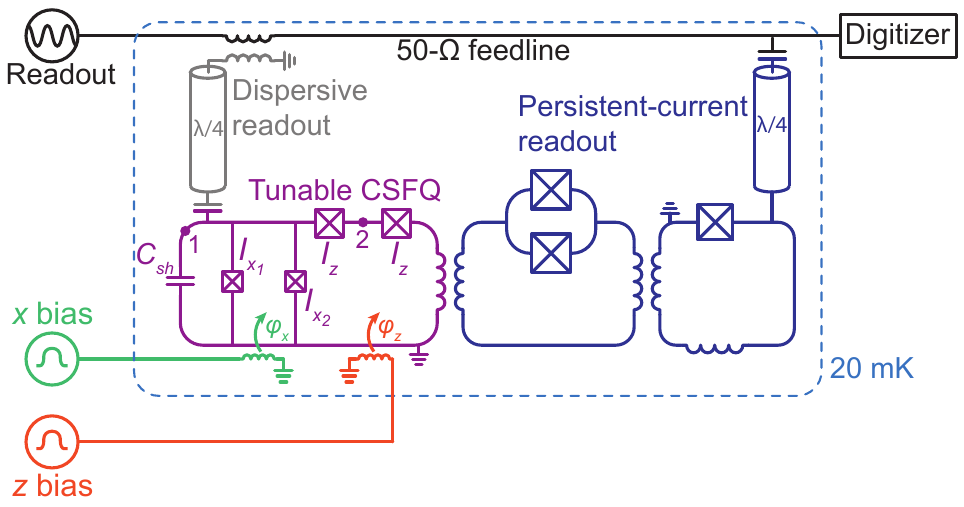}
    \end{center}
    \caption{
	Simplified schematic of the experimental set up.
	The 4-junction CSFQ (purple) is controlled via two bias lines that thread $x$ (green) and $z$ (orange) fluxes into their corresponding loops.
	The flux waveforms have 1-ns time resolution.
	The qubit is coupled to a dispersive readout resonator (gray) and to a persistent-current readout circuit (dark blue).
	The latter measures the direction of the circulating current in the qubit $z$ loop.
	Note that the CSFQ floats in the physical device; the displayed ground defines the zero point for the circuit node parameters in the Hamiltonian derivation.
	Details of the persistent-current readout can be found in~\cite{Grover2020}.
	Relevant device parameters are discussed in Methods.
    }\label{fig:device}
\end{figure}

In this work we demonstrate an alternative and complementary approach with a CSFQ: using dispersive measurements to quantify the asymmetry in the qubit junctions, we use our component-level circuit model (Fig.~\ref{fig:device}) to devise corrected annealing paths that dynamically cancel the nonlinear crosstalk effect.
Our approach is designed for high-coherence CSFQs, since they use fewer superconducting loops and bias lines to reduce flux noise and coupling to the environment.
Additionally, this is the natural choice for high-control CSFQs that are capable of implementing customized annealing schedules.
We use this approach to demonstrate a twofold reduction in the ``s-curve" transition width between the qubit wells as a function of applied tilt bias, without adding any additional circuit elements.
To confirm the validity of the multi-level circuit model used for anneal path corrections, we anneal the qubit through small gaps and transfer the population to higher excited states.
We then use our circuit model, which is fit to independently measured spectroscopy data, to accurately predict the population exchanges, and use the adiabatic master equation to qualitatively explain the observed open system effects.

\section{Results}

\begin{figure*}[t]
	\begin{center}
		\includegraphics[width=0.95\textwidth]{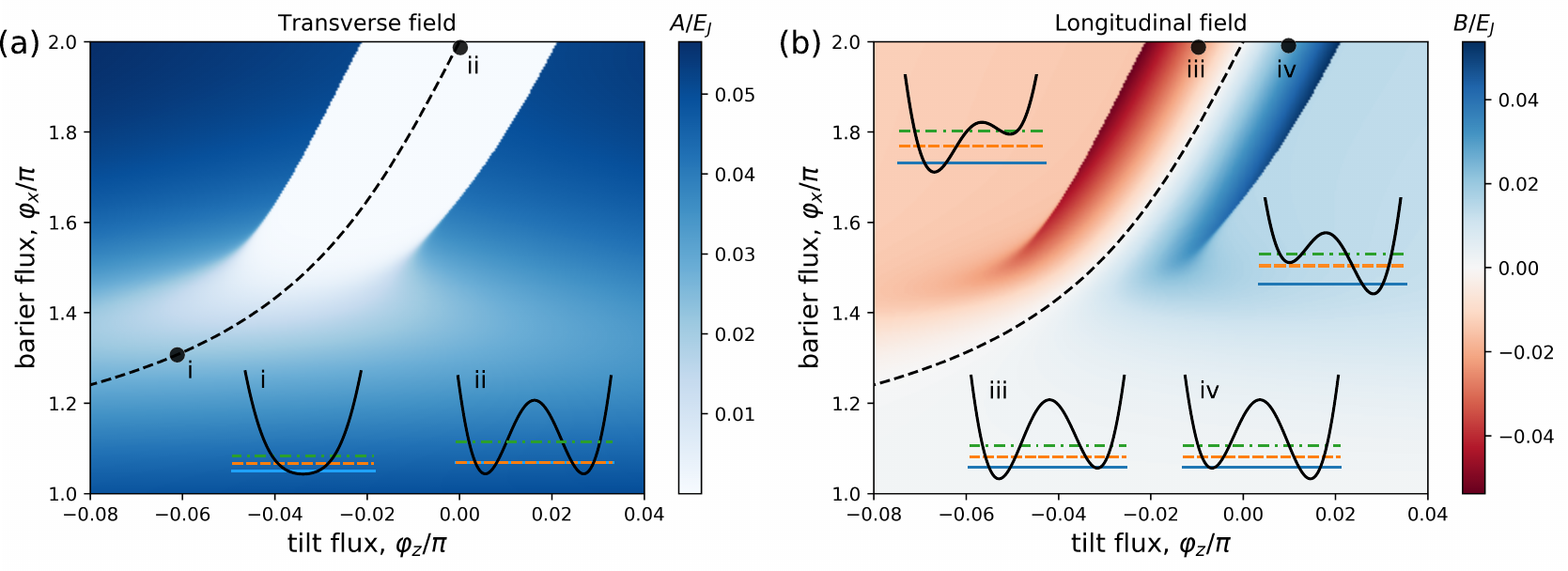}
    \end{center}
    \caption{
    Numerically calculated transverse field Ising Hamiltonian coefficients [Eq.~\eqref{eq:H-ann}] as a function of flux biases.
	(a) Transverse field coefficient $A(t)$ for $\sigma_x$. 
	Filled circle (i) marks the bias values corresponding to $A(0)$ for which there is only a single well (inset i) and the transverse field is large, and (ii) marks the bias values  at the degeneracy point $\varphi_\text{z}=0$ corresponding to $A(t_f)$ for which the barrier is high (inset ii) and the tunneling between the symmetric wells is suppressed; the lowest two levels (solid blue and dashed orange lines) are degenerate.
	An annealing path corresponds to moving from point (i) to point (ii), during which the barrier is raised.
	The dashed black lines show the location of the qubit's degeneracy point (minimum gap).
	The reason for the skewed shape is the asymmetry effect described in the text.
	(b) Longitudinal field coefficient $B(t)$ for $\sigma_z$.
	Filled circle (iii) marks the bias values for which the double well potential is tilted to the left (inset iii) and the longitudinal field is negative, and (iv) marks the values for which the double well potential is tilted to the right (inset iv) and the longitudinal field is positive.
	For very large tilt values, outside of the boundaries defined by the sharp color transition, the first two eigenenergies are both localized in the same well, as illustrated by the insets inside panel (b).
	When this happens the Ising mapping no longer applies and the circuit cannot be used as a flux qubit.
	For our parameters, $E_\text{J}/2\pi \approx 100$~GHz.}
\label{fig:panel}
\end{figure*}

\subsection{System and model}
The experimental setup is depicted in Fig.~\ref{fig:device}.
We use a four-junction CSFQ~\cite{Yan2016}, controlled with two flux bias lines that thread external fluxes into the loops of the qubit.
The CSFQ is coupled to a dispersive readout resonator at $\omega_\text{r}/2\pi=$ 7.1876~GHz, which is used to calibrate the linear crosstalk between the $x$- and $z$-flux bias lines~\cite{Novikov2018}, and to send microwave pulses to the qubit.
Our device is also equipped with a persistent current readout that measures the direction of the circulating current in the large $z$-loop (see Supplementary Note 2).

The Hamiltonian of the CSFQ circuit can be written as
\begin{align}
\label{eq:H}
	H =& \frac{e^2}{2C_\text{sh}+(4\alpha+1)C_\text{z}}(2\hat{n}_1+\hat{n}_2)^2 + \frac{e^2}{C_\text{z}}\hat{n}_2^2 \nonumber \\
	 -& 2 I_\text{z} \frac{\Phi_0}{2\pi} \cos(\hat{\varphi}_2-\hat{\varphi}_1/2)\cos(\hat{\varphi}_1/2-\varphi_\text{z}/2)  \\
	 -& 2 \alpha I_\text{z} \frac{\Phi_0}{2\pi} \cos(\varphi_\text{x}/2)\sqrt{1+\tan^2(\varphi_\text{d})}\cos(\hat{\varphi}_1 - \varphi_\text{d}), \nonumber
\end{align}
where the operators $\hat{\varphi}_{k}$ and $\hat{n}_{k}$ are, respectively, the superconducting phase and number of Cooper pairs at circuit nodes $k=1,2$, satisfying the commutation relation $[\hat{\varphi}_k, \hat{n}_l] = i\delta_{kl}$ (see Methods for derivation).
Note that phase and flux are related through $\varphi_{i}=2\pi\Phi_{i}/\Phi_0$. Here $\Phi_0$ is the magnetic flux quantum, and $\varphi_\text{x}$ and $\varphi_\text{z}$ are the barrier and tilt, respectively, also referred to as the $x$ and $z$ (flux) bias (see Fig.~\ref{fig:panel}).
$C_\text{sh}$ is the shunt capacitance, $C_\text{z}$ is the capacitance of each of the two $z$-loop junctions whose critical currents are $I_\text{z}$.
The $x$-loop junctions are on average $\alpha$ times smaller than the $z$-loop junctions, such that $(I_\text{x1}+I_\text{x2})/2=\alpha I_\text{z}$ and $(C_\text{x1}+C_\text{x2})/2=\alpha C_\text{z}$, where $C_\text{xi}$ is the capacitance of the $i$th $x$-loop junction.
A central role is played in our experiments by the asymmetry between the two $x$-loop junctions.
We define an asymmetry parameter as $d \equiv (I_\text{x1}-I_\text{x2})/(I_\text{x1}+I_\text{x2})$, with its corresponding phase shift
\begin{equation}
\label{eq:phi_d}
\varphi_\text{d} = \arctan[d\tan(\varphi_\text{x}/2)] .
\end{equation}
Note that the asymmetry parameter $d$ is independent of the $x$-bias and is a property of fabricated circuits, but the asymmetry induced phase shift of Eq.~\eqref{eq:phi_d} depends on it.
Eq.~\eqref{eq:H} shows that the asymmetry of the $x$-loop junctions rescales the total current through them by $\sqrt{1+\tan^2(\varphi_\text{d})}$ and also shifts the $z$-loop bias as $\varphi_\text{z} \mapsto \varphi_\text{z} - \varphi_\text{d}$.
This $\varphi_\text{x}$-dependent shift of the $z$-bias is a nonlinear quantum crosstalk induced by the junction asymmetry, and must be taken into account when operating the CSFQ in annealing protocols.

We note that the standard QA Hamiltonian of a single qubit is obtained from the circuit Hamiltonian~\eqref{eq:H} by retaining only the lowest two energy eigenstates (see Supplementary Note 3), which yields:
\begin{equation}
\label{eq:H-ann}
H_q(t) = A(t) \sigma_x + B(t) \sigma_z ,
\end{equation}
where $\sigma_{x}$ and $\sigma_{z}$ are the Pauli matrices representing the transverse and longitudinal fields, respectively, and $A(t)$ and $B(t)$ are the time-dependent annealing schedules, with $t\in[0,t_f]$.
Time-dependent paths in flux space control the transverse and longitudinal fields of the annealing schedule.

In general, such two-level reduction works as long as non-adiabatic transitions to states outside the chosen computational subspace can be neglected.
Additionally, for flux qubits we require the lowest two eigenstates to have support in both wells of the potential, i.e., not be localized in the same well.
This imposes an upper bound on $|\varphi_\text{z}|$, as illustrated in Fig.~\ref{fig:panel}(b).
In Supplementary Note 3 we identify this bound on the $z$-bias and provide expressions for the annealing schedules in terms of the circuit Hamiltonian parameters.

\subsection{Asymmetry measurement}
We measure $d$ by noting that the qubit's minimum gap occurs at $\varphi_\text{z}^\text{min}= \varphi_\text{d}$ when $ \pi \leq\varphi_\text{x} \leq 3\pi$ (see Supplementary Note 1).
As illustrated in Fig.~\ref{fig:d}, we scan the $x$ and $z$-biases around the qubit's minimum gap and measure the demodulated signal of the dispersive readout resonator, corresponding to an energy eigenbasis measurement of $H$. 
For a fixed $\varphi_\text{x}$, the dispersive readout signal is symmetric as a function of $\varphi_\text{z}$ relative to the minimum gap position (symmetry point).
We fit a Gaussian to the readout signal along $\varphi_\text{z}$ to extract this position, and repeat for all values of $\varphi_\text{x}$ (filled green circles in Fig.~\ref{fig:d}).
We then use Eq.~\eqref{eq:phi_d} to fit this data to $\varphi_\text{z}^\text{min}(d,\varphi_\text{x})$ (dashed line in Fig.~\ref{fig:d}) to extract the asymmetry parameter, albeit with offsets on both fluxes that are fitted as well to account for flux drifts and/or offsets.
We obtain $d = 0.102 \pm 0.005$, where the value was determined by systematically varying the fitting regions and using resampling to compute the $1\sigma$ confidence interval.

Note that in our system, similar dispersive measurements of the qubit as a function of $x$ and $z$-biases are performed for linear crosstalk calibration of the flux bias lines \cite{Novikov2018}, and the asymmetry extraction discussed above is simply a different post-measurement analysis of the same data.
The junction asymmetry is a property of fabricated circuits that is local to each individual circuit element, therefore our asymmetry measurement procedure is easily extensible to multi-qubit systems, where it is again similar to local linear crosstalk calibration.

\begin{figure}[t]
	\begin{center}
		\includegraphics[width=0.95\columnwidth]{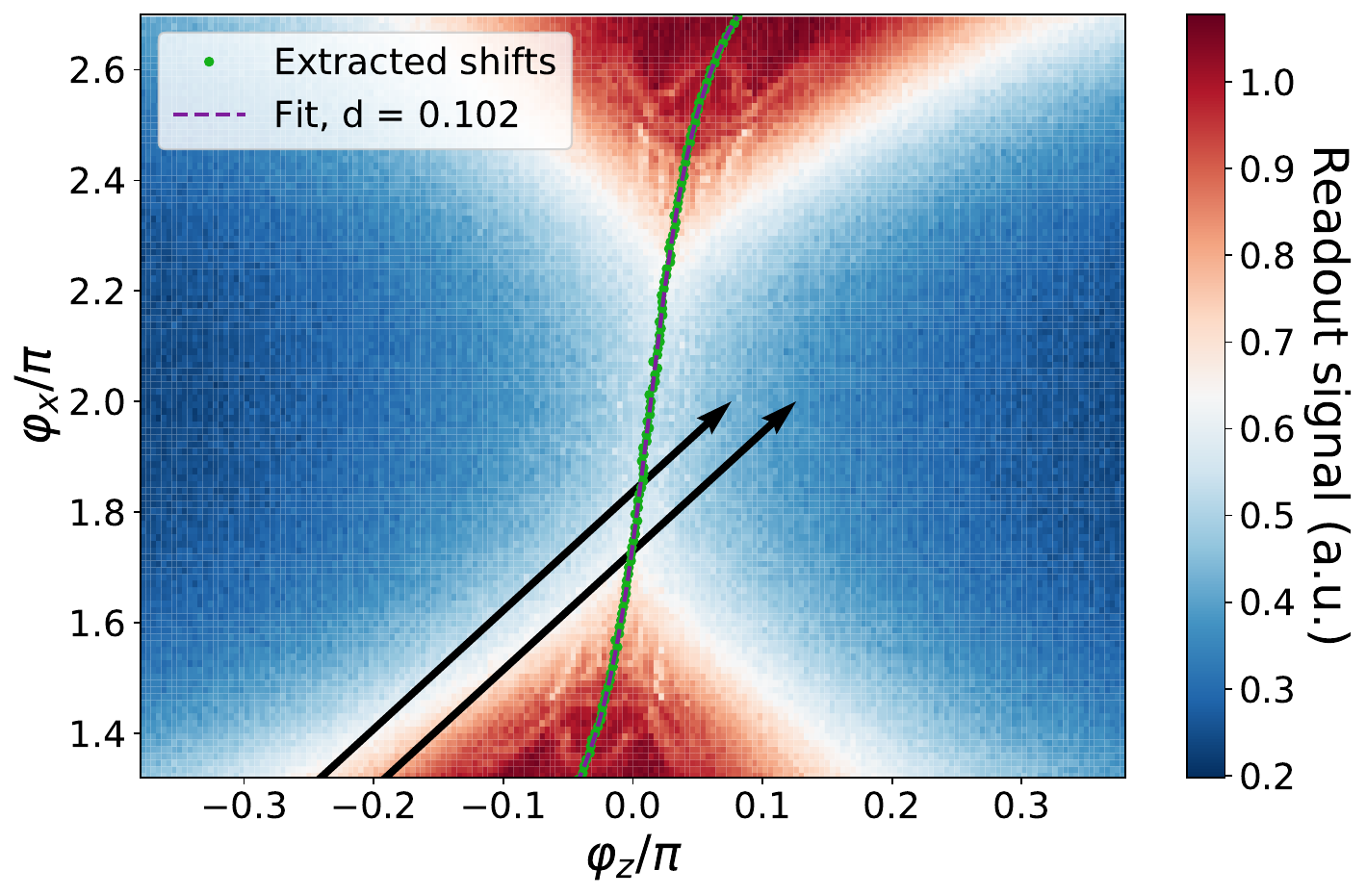}
    \end{center}
    \caption{
    Extraction of junction asymmetry from the experimentally measured dispersive resonator response, corresponding to the $0\leftrightarrow1$ transition frequency of the CSFQ.
	The green circles are the extracted symmetry-point shifts, and the dashed purple line denotes the fit to theory.
	The slanted arrows represent annealing paths used to probe the multilevel structure of the CSFQ, corresponding to two of the data-points in Fig.~\ref{fig:crossing}.
    }\label{fig:d}
\end{figure}

\subsection{S-curve width reduction via annealing path control}
To characterize our device for use in QA experiments, we perform a so-called ``s-curve" measurement~\cite{Chiorescu2003, Segall2003, Crankshaw2004, Oliver2005, Valenzuela2006, Berkley2010, Harris2010, Kwon2011, Quintana2017} on our CSFQ.
This is a single-qubit annealing experiment, where the CSFQ starts in the single-well regime [$A(0)\gg B(0)$] with a variable initial tilt $\varphi_\text{z}$, then the barrier $\varphi_\text{x}$ is raised (at fixed $\varphi_\text{z}$) to put the qubit in a tilted double-well regime, with negligible tunneling between the two wells.
This is illustrated in Fig.~\ref{fig:panel}(a) (also see Supplementary Note 3).
Finally, a persistent current measurement is performed to determine which of the wells is occupied, corresponding to a computational basis measurement of $\sigma_z$.
Ideally, the s-curve would be a step function.
In actuality, one obtains a curve that resembles an S shape with a characteristic width for transitioning between left and right circulating currents at the degeneracy point. 
The width $w$ can be found by fitting the right-well population $P$ to a phenomenological model~\cite{Harris2010}:
\begin{equation}
	P = \frac{1}{2}\left[ 1 + \tanh\left( \frac{\varphi_\text{z} - \varphi_\text{z0}}{w} \right) \right].
\end{equation}
The width, which should be minimized, depends on the rate at which the barrier is raised, thermalization between the states in left and right wells, and flux noise in the tilt bias near the minimum gap~\cite{Chiorescu2003, Segall2003, Crankshaw2004, Oliver2005, Valenzuela2006, Berkley2010, Harris2010, Kwon2011, Quintana2017phd}.
In an annealing process, minimizing the s-curve width improves performance by increasing the qubit's sensitivity to other qubits it is coupled to, and increases the dynamic range of couplers by making it easier to induce a detectable shift in the qubit's state.

\begin{figure}[t]
	\begin{center}
		\includegraphics[width=0.9\columnwidth]{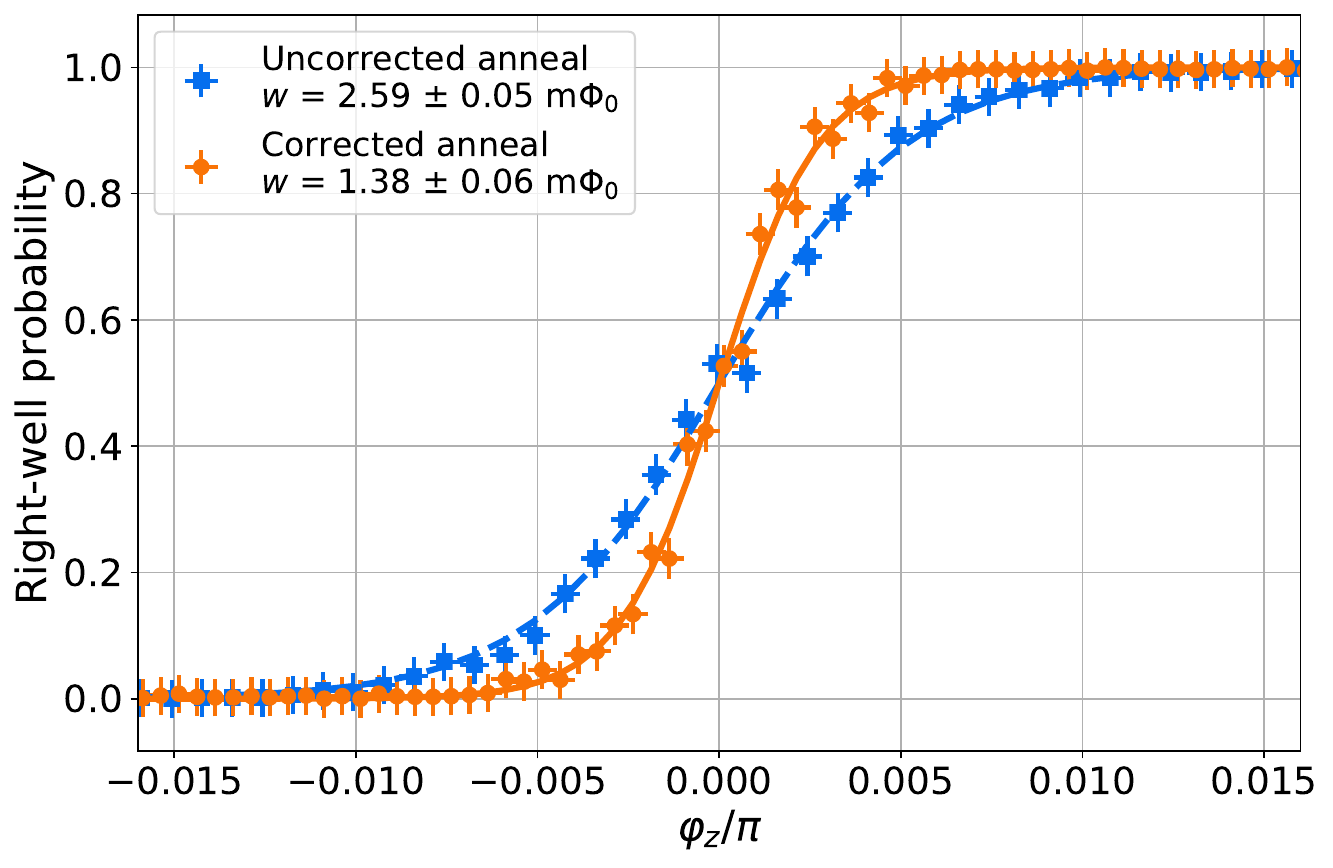}
    \end{center}
	\caption{
		S-curve without (blue) and with (orange) annealing-path correction.
		In the former each datapoint is obtained by sweeping only $\varphi_\text{x}$ at fixed $\varphi_\text{z}$.
		In the latter, we added $\varphi_\text{d}(\varphi_\text{x})$ to $\varphi_\text{z}$. 
		Both anneals occur in 20 ns.
		The uncorrected anneal (squares) results in a width of $2.58 \pm 0.05~\mathrm{m}\Phi_0$ (fit, dashed line).
		Applying the correction (circles) narrows the s-curve by nearly $50\%$ to $1.38 \pm 0.06~\mathrm{m}\Phi_0$ (fit, solid line).
		While both sets of data have been shifted and centered for ease of comparison, the corrected anneal should center the curve around the degeneracy point, which can be used to calibrate offsets in the $z$ bias.
		Error bars show standard deviation and are calculated from binomial counting statistics, AWG voltage resolution, and quasistatic noise.
	}
	\label{fig:scurve}
\end{figure}

Nonlinear crosstalk also acts to increase $w$.
Namely, when the $x$-bias (barrier) is tuned during an s-curve measurement, the junction asymmetry causes an extra tilt of the potential if the $z$-bias is kept constant.
This shifts the center of the s-curve away from the degeneracy point, and broadens its width; see the blue dashed curve in Fig.~\ref{fig:scurve}.
To cancel this effect, we correct the annealing path with respect to the junction asymmetry by applying an additive $z$-bias correction of $+\varphi_\text{d}$ [Eq.~\eqref{eq:phi_d}], to undo the asymmetry-induced shift of $\varphi_\text{z} \mapsto \varphi_\text{z}-\varphi_\text{d}$ (See Supplementary Note 7 for correction pulse details).
This amounts to a nonlinear annealing path in the $(\varphi_\text{z},\varphi_\text{x})$ plane.
We note that for inductively coupled qubits in a multi-qubit annealing setting, where qubit-qubit interactions are mediated by their persistent currents, in addition to the correction of the $z$-bias due to asymmetry, the $x$-bias should also be adjusted to undo the asymmetry-induced rescaling of the persistent current by $\sqrt{1+\tan^2(\varphi_\text{d})}$.

Our first key result is a reduction of the s-curve width by nearly $50\%$ when comparing the standard (fixed $\varphi_\text{z}$) s-curve protocol to our protocol that corrects for the asymmetry-induced nonlinear crosstalk, as shown by the orange solid line in Fig.~\ref{fig:scurve}.
This substantial improvement is made possible by two key capabilities: first, the independent extraction of the asymmetry parameter $d$ via dispersive measurement, and second the independent individual control we have over the flux biases, which enables an accurate traversal of the optimal, nonlinear annealing path shown in Fig.~\ref{fig:d}.
Note that asymmetry extraction, and in general crosstalk calibration, may alternatively be performed using the available tunable resonator used for the persistent current readout, eliminating the need for the dispersive resonator and reducing system complexity when scaling up the system.
This will be the subject of a future study.

\begin{figure*}[t]
	\begin{center}
		\subfigure{\includegraphics[width=0.329\textwidth]{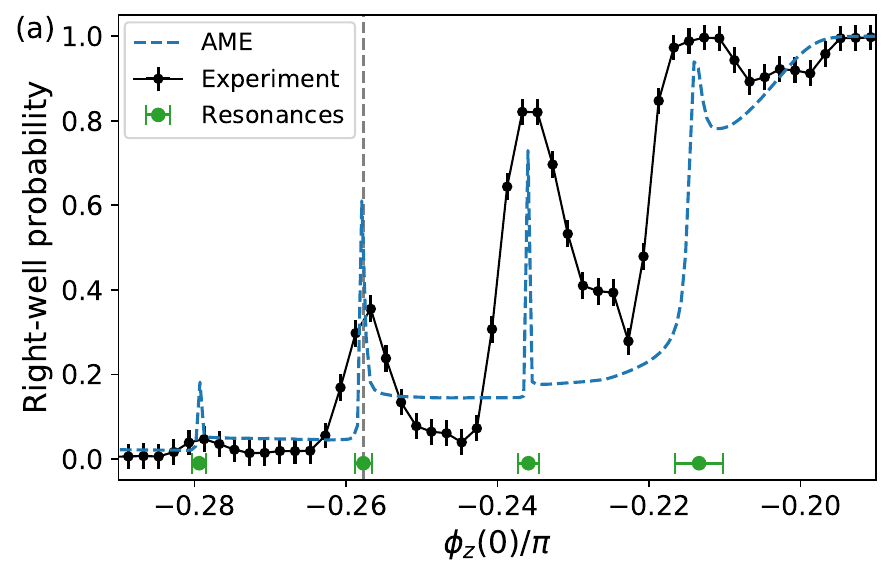}\label{fig:crossing-a}}
		\subfigure{\includegraphics[width=0.329\textwidth]{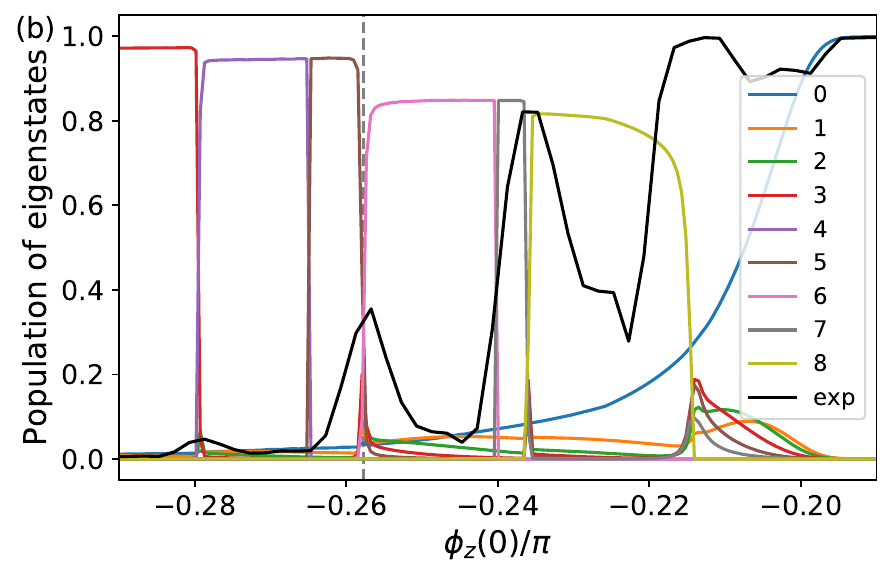}\label{fig:crossing-b}}
		\subfigure{\includegraphics[width=0.329\textwidth]{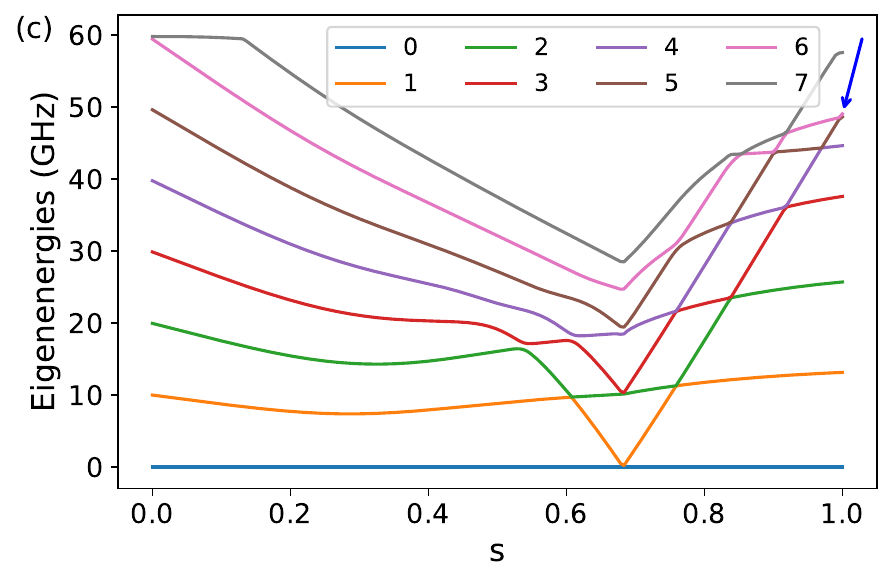}\label{fig:crossing-c}}
    \end{center}
	\caption{Resonant features of quantum multi-level structure of the CSFQ.
		(a) Persistent current readout giving the right-well probability as a function of the initial tilt bias, for a linear anneal in both $\varphi_\text{x}$ and $\varphi_\text{z}$, as illustrated by the slanted arrows in Fig.~\ref{fig:d}.
		Tilt bias anneal amplitude is $\text{amp}=0.326\pi$, i.e., $\varphi_\text{z}(t) = \varphi_\text{z}(0) + \text{amp}\times(t/t_f)$, where $t_f=60$~ns is the anneal time.
		The black solid line shows the experimental result, and the dashed blue line is the the AME result for the CSFQ circuit with parameters extracted from spectroscopy (see text).
		Experimental error bars show standard deviation and are calculated as in Fig.~\ref{fig:scurve}.
		(b) Population of CSFQ circuit eigenstates (indicated by the numbers in the legend, with $0$ being the ground state) at the end of each anneal, along with the experimental result (exp) also shown in (a).
		Only avoided level crossings lead to observable features in the persistent-current readout, indicated by the green circles in (a).
				(c) An example of the CSFQ spectrum vs normalized anneal time $s=t/t_f$, for an initial tilt bias corresponding to the grey vertical dashed line in panels (a) and (b).
		The blue arrow marks the avoided level crossing between levels $5$ and $6$ at the end of the anneal, which corresponds to the population exchange between these two levels in panel (b), and the corresponding experimental resonant feature.
		Cascaded level crossings that transfer the population are visible throughout the anneal.
	}\label{fig:crossing}
\end{figure*}

\subsection{Signatures of level crossing}
So far we have used the circuit model of Eq.~\eqref{eq:H} to measure and analyze the effect of junction asymmetry, and to find annealing paths that correct for the asymmetry induced nonlinear crosstalk.
In this section, we validate and justify our circuit model by fitting it to spectroscopy data and using the fitted model to investigate and explain the multilevel structure of our CSFQ circuit.

We probe the dispersive resonator while driving the qubit to perform standard two-tone spectroscopy~\cite{Schuster2005}, varying $\varphi_\text{z}$ and $\varphi_\text{x}$ to change the qubit frequency.
Using a high qubit drive power allows us to extract the two lowest transition frequencies $\omega_{01}$ and $\omega_{02}$ of the circuit (see Supplementary Note 1 for the data).
We then find the circuit parameters of our model by fitting the two lowest transition frequencies of the Hamiltonian~\eqref{eq:H} to spectroscopy measurements (see Table~\ref{tab:fit} fit values), with strong agreement between the fitted model and experimental data (see Methods and Supplementary Note 1 for details).
This gives us a fitted circuit model that we can use to predict other behaviors of our qubit, as we discuss below.

To investigate the multilevel circuit model, we perform a modified s-curve measurement where in addition to raising the barrier $\varphi_\text{x}$, we also linearly increase the tilt $\varphi_\text{z}$ during each anneal, and repeat for different initial values $\varphi_\text{z}(0)$ (illustrated by the slanted lines in Fig.~\ref{fig:d}).
During such anneals, the gap of the qubit closes and the population is diabatically transferred to higher qubit energy levels.
The persistent current measurement results obtained at the end of each anneal are shown by the solid black line in Fig.~\ref{fig:crossing-a}.
The overall behavior resembles the s-curve of Fig.~\ref{fig:scurve}, but now exhibits a much wider transition domain, accompanied by multiple sharp features~\cite{Crankshaw2004}.
Note that this also shows that a linear correction to the tilt bias is insufficient for mitigating the asymmetry-induced crosstalk that broadens the s-curve width.
We proceed to establish that these features represent resonances between the quantized higher energy levels of the CSFQ circuit.

To explain the resonances [peak features in Fig.~\ref{fig:crossing-a}], we theoretically calculate the spectrum of the Hamiltonian~\eqref{eq:H} along the same annealing paths as implemented experimentally, using the aforementioned independently extracted circuit parameters.
The CSFQ is initially in its ground state, but as shown in Fig.~\ref{fig:crossing-c} a cascade of avoided and unavoided (actual) level crossings takes place during the anneal, so that the population is diabatically transferred to higher energy levels.
The initial tilt bias $\varphi_\text{z}(0)$ determines the most-populated level at the end of each anneal, as can be seen in Fig.~\ref{fig:crossing-b}.
For a given initial tilt $\varphi_\text{z}(0)$, an experimental peak is observed if an avoided level crossing occurs at the end of that anneal, but no such peak is observed if an anneal ends with an unavoided level crossing.
The green circles in Fig.~\ref{fig:crossing-a} correspond to those $\varphi_\text{z}(0)$ values for which the anneal ends with an avoided crossing, calculated using extracted circuit parameters and Eq.~\eqref{eq:H}, and accurately predict the locations of the experimental peaks.
The error bars are due to uncertainty in the fitted circuit parameters.
We emphasize that the theoretical peak locations in this experiment [the green circles in Fig.~\ref{fig:crossing-a}] are calculated using circuit parameters that are extracted via independent spectroscopy measurement.
This involves only a static calculation of the energy spectrum of the circuit, without any dynamics.

This population transfer mechanism explains the peak features seen in Fig.~\ref{fig:crossing-a}: as we vary $\varphi_\text{z}(0)$, a previously unoccupied eigenstate crosses with the occupied eigenstate and suddenly acquires its population (a resonance).
Consequently there is a sudden change in the result of the persistent-current readout, because the right-well population measured at the end of each anneal depends on the population in each eigenstate, the persistent-current value associated with that eigenstate, and the persistent-current readout resolution.
Only avoided level crossings yield a persistent-current feature that is observable in the experiment, since for actual level crossings the population is completely transferred to other eigenstates and the total persistent current of the CSFQ does not change enough to yield an observable feature.

To observe the population transfer between eigenstates, and also to account for open system effects, we simulate the dynamics of the circuit described by Eq.~\eqref{eq:H} using the adiabatic master equation (AME)~\cite{Albash2012} (see Methods for details).
We use the same annealing paths that were implemented in our experiments and assume an Ohmic bath at $10$~mK that is weakly coupled to the system, with a high-frequency cutoff at $\omega_\text{c}/2\pi = 15$~GHz.
We also add a $2$~ns idle time at the end of each anneal to mimic the effect of delay before the persistent-current readout in the experiment, which allows for relaxation (without this delay the features manifest as plateaus; see Supplementary Note 6).
The result is the blue dashed line in Fig.~\ref{fig:crossing-a} that accurately predicts the locations of the resonances and qualitatively captures their behavior, namely the existence of peaks at resonances and the relative magnitudes of these peaks.
The eigenstate occupations at the end of each anneal are also plotted in Fig.~\ref{fig:crossing-b}, showing population exchange between circuit levels at energy crossings as expected.

Although the AME simulations with independently measured circuit parameters show good qualitative agreement with experiment and confirm the multi-level cascaded population exchange between the states, they yield narrower features than the experimental results shown in Fig.~\ref{fig:crossing-a}.
Similar differences between AME simulations and experimental features were observed before~\cite{Albash2015b}, which is not surprising given that the AME with an Ohmic bath discards low-frequency noise, known to be a dominant source in superconducting qubits~\cite{Bialczak2007,Yan2012,Anton2013,Quintana2017} (see Methods).

\section{Discussion}
We have demonstrated a hardware-level quantum control approach to overcoming the nonlinear crosstalk between control fluxes arising from fabrication variation of Josephson junctions in flux qubits.
Our approach implements the necessary nonlinear anneal-path correction, while avoiding the introduction of additional control lines or circuit elements.
We have used this to demonstrate a 50\% reduction in the s-curve width for our qubits, and also showed that a linear correction to the tilt bias is insufficient for mitigating the asymmetry induced s-curve broadening.
Note that the s-curve consists of a series of single-qubit annealing experiments, whose transition width would vanish (a step function) in the limit where every anneal is perfectly successful.
However, in practice there is a transition width that depends on the rate at which the barrier is raised, thermalization between the states in left and right wells, flux noise in the tilt bias specifically around the minimum gap~\cite{Quintana2017phd}, and as shown in this work the proper choice of flux controls (i.e., anneal paths).
Therefore the width is a characteristic of the noise environment of the qubit, as well as system operation and control fidelity, assuming the anneals are slow enough to avoid broadening due to nonadiabatic effects.

One can associate an effective temperature to the qubit's s-curve width by multiplying it by the persistent current of the qubit at the end of the anneal to get
\begin{equation}
	T_\text{eff} = \frac{w I_\text{p}}{k_B},
\end{equation}
which we use to compare the s-curve width between multiple platforms and qubit designs.
Note that near the degeneracy point, $wI_\text{p}$ is the effective longitudinal field ($\sigma_z$ coefficient) in the Ising spin model of qubits.
Since both $w$ and $I_\text{p}$ should be minimized (recall that slow flux noise degrades the energy relaxation and the coherence time, scaling roughly as $1/I_\text{p}^2$ and $1/I_\text{p}$, respectively), a smaller $T_\text{eff}$ is preferable.
The relevant dimensionless quantity is $T_\text{eff}$ scaled by the dilution fridge temperature, $T_\text{fridge}$.
Table~\ref{tab:width} shows a summary of $T_\text{eff}/T_\text{fridge}$ values across different flux qubit designs.
\begin{table*}[ht]
	\begin{center}
    \begin{ruledtabular}
    \begin{tabular}{ | c | c | c | c | c | c |}
    Group & $w\, (\mu\Phi_0)$ & $I_\text{p}\, (\mu \textrm{A})$ & $T_\text{fridge}\, (\textrm{mK})$ & $T_\text{eff}\, (\textrm{mK}) $ & $T_\text{eff} / T_\text{fridge}$ \\ \hline
    D-Wave & 45 & 2.6~\cite{Harris2010} & 8 & 17.5 & 2.2 \\ \hline
    Google & 100~\cite{Quintana2017phd} & 0.87~\cite{Quintana2017phd} & 10~\cite{Quintana2017} & 13 & 1.3 \\ \hline
	QEO (this work) & 1400 & 0.17 & 20 & 35.7 & 1.8 \\ \hline
    QEO (improved) & 760 & 0.17 & 15 & 19.4 & 1.3
    \end{tabular}
    \end{ruledtabular}
    \caption{
    Comparison of s-curve widths across experimental groups, where QEO stands for the Quantum Enhanced Optimization collaboration.
    The raw width is multiplied by the qubit persistent current, $I_\text{p}$, to convert it to an effective temperature, $T_\text{eff}$.
    This is then divided by the dilution refrigerator operating temperature, $T_\text{fridge}$, to create a dimensionless quantity for cross-platform comparison.
    The improved QEO width was measured on a colder fridge with better line filtering, and will be discussed in a future publication.
    The D-Wave s-curve width and dilution refrigerator temperature are reported from private communications with their researchers.
    }\label{tab:width}
    	\end{center}
\end{table*}

The mitigation of the asymmetry-induced nonlinear crosstalk reduces the s-curve width, and it does so by increasing the success probability of its single qubit anneals.
In a broader sense, asymmetry-induced crosstalk correction enhances the system operation fidelity, which yields improvement in success probability of annealing protocols involving multiple qubits.
For this reason the results presented here are an important step on the path towards achieving high-fidelity annealing operation of high-coherence and high-control flux qubits, a critical enabling capability in constructing quantum annealers exhibiting a quantum advantage.
Improvements in chip designs, fridge line filtering, and pulse distortion calibration can lead to more accurate control of quantum annealing systems, which will be pursued in future work.

\section{Methods}

\section{Derivation of the CSFQ Hamiltonian}

In this section we give a detailed derivation of the CSFQ Hamiltonian of Eq.~\eqref{eq:H}.
For clarity and completeness, we repeat some of the details given there.

The capacitively shunted flux qubit (CSFQ) has two superconducting loops, each terminated with two junctions, shunted with a large capacitance (Fig.~\ref{fig:device}).
The $x$-loop is threaded with an external flux $\Phi_x=\Phi_0\varphi_\text{x}/2\pi$, which controls the height of the barrier in the double well potential.
The larger $z$-loop is threaded with $\Phi_z=\Phi_0\varphi_\text{z}/2\pi$, which tilts the double-well potential.
In our experiment, the qubit is coupled to a dispersive readout resonator, and also has persistent-current readout that can measure the direction of the circulating current in the $z$-loop.
The nodes $1$ and $2$ used for derivation of the Hamiltonian of this circuit are marked with filled circles in Fig.~\ref{fig:device}.
Here for simplicity we ignore the qubit's inductance in the Hamiltonian derivation, knowing that its contribution to the energy levels of the qubit is negligible.

The capacitance matrix of the above circuit can be written as
\begin{align}
	\mathbf{C} = & 
	\begin{pmatrix}
		C_\text{sh} + C_\text{x1} + C_\text{x2} + C_\text{z} & -C_\text{z} \\
		-C_\text{z} & C_\text{z} + C_\text{z} & \\
	\end{pmatrix} \nonumber \\
	= &
	\begin{pmatrix}
		C_\text{sh} + (2\alpha+1)C_\text{z} & -C_\text{z} \\
		-C_\text{z} & 2C_\text{z} & \\
	\end{pmatrix},
\end{align}
where $C_\text{sh}$ is the shunt capacitance, and $C_\text{z}$ is the capacitance of the $z$-loop junction that has a critical current of $I_z$. 
The $x$-loop junctions are on average $\alpha$ times smaller than the $z$-loop junctions, such that $(I_\text{x1}+I_\text{x2})/2=\alpha I_z$ and $(C_\text{x1}+C_\text{x2})/2=\alpha C_\text{z}$, where $C_{xi}$ and $I_{xi}$ are the capacitance and critical current of the $i$th $x$-loop junction respectively.
The kinetic energy of the circuit is then
\begin{align}
	K_\text{2D} = & \frac{1}{2}(2e)^2 \vec{n}^T\cdot\mathbf{C}^{-1}\cdot\vec{n} \nonumber \\
	= & \frac{e^2}{C_\text{z}}\hat{n}_2^2 + \frac{e^2}{2C_\text{sh}+(4\alpha+1)C_\text{z}}(2\hat{n}_1+\hat{n}_2)^2,
\end{align}
where $\vec{n}=(n_1, n_2)$ is a column vector of the number of Cooper pairs at each node.

To write the potential energy, we choose a gauge that splits (symmetrizes) the control fluxes over both of its junctions, to get:
\begin{align}
	U_\text{2D} = -\frac{\Phi_0}{2\pi}\big[ & I_\text{x1}\cos(\varphi_1 - \varphi_\text{x}/2) + I_\text{x2}\cos(\varphi_1 + \varphi_\text{x}/2) \nonumber \\
	& + I_z\cos(\varphi_1 - \varphi_2 - \varphi_\text{z}/2) \nonumber \\
	& + I_z\cos(\varphi_2 - \varphi_\text{z}/2) \big],
\end{align}
where $\hat{\varphi}_1$ and $\hat{\varphi}_2$ are the superconducting phases at nodes $1$ and $2$, satisfying commutation relation $[\hat{\varphi}_k, \hat{n}_l] = i\delta_{kl}$.
Note that phase and flux are related through $\varphi_i=2\pi\Phi_i/\Phi_0$, where $\Phi_0$ is the magnetic flux quantum.
By defining the qubit asymmetry parameter as $d \equiv (I_\text{x1}-I_\text{x2})/(I_\text{x1}+I_\text{x2})$ and its corresponding phase shift as 
\begin{equation}\label{eq:phi_ds}
	\tan(\varphi_\text{d}) \equiv d\tan(\varphi_\text{x}/2) ,
\end{equation}
after some algebra we can simplify the potential energy as:
\begin{align}\label{eq:U_2D}
	 U_\text{2D} &= - 2 I_z \frac{\Phi_0}{2\pi} \cos(\hat{\varphi}_2-\hat{\varphi}_1/2)\cos(\hat{\varphi}_1/2-\varphi_\text{z}/2) \nonumber \\
	 -& 2 \alpha I_z \frac{\Phi_0}{2\pi} \cos(\varphi_\text{x}/2)\sqrt{1+\tan^2(\varphi_\text{d})}\cos(\hat{\varphi}_1 - \varphi_\text{d}).
\end{align}
The Hamiltonian of the CSFQ circuit can then be written as
\begin{align}\label{eq:H_2D}
	H_\text{2D} &=  K_\text{2D} + U_\text{2D} \\
	&= \frac{e^2}{C_\text{z}}\hat{n}_2^2 + \frac{e^2}{2C_\text{sh}+(4\alpha+1)C_\text{z}}(2\hat{n}_1+\hat{n}_2)^2 \nonumber \\
	 &- 2 I_z \frac{\Phi_0}{2\pi} \cos(\hat{\varphi}_2-\hat{\varphi}_1/2)\cos(\hat{\varphi}_1/2-\varphi_\text{z}/2) \nonumber \\
	 &- 2 \alpha I_z \frac{\Phi_0}{2\pi} \cos(\varphi_\text{x}/2)\sqrt{1+\tan^2(\varphi_\text{d})}\cos(\hat{\varphi}_1 - \varphi_\text{d}). \nonumber
\end{align}

We can transform the coordinates in \eqref{eq:H_2D} to diagonalize the kinetic part of the Hamiltonian.
This will allow us to identify and separate fast and slow degrees of freedom in our Hamiltonian and further simplify our circuit model.
The coordinate transformation that satisfies the commutation relations is 
\begin{subequations}
\begin{align}\label{eq:trans}
	\begin{pmatrix}
		n_1' \\
		n_2'
	\end{pmatrix}
	&=
	\begin{pmatrix}
		0 & 1 \\
		2 & 1
	\end{pmatrix}
	\begin{pmatrix}
		n_1 \\
		n_2
	\end{pmatrix}
	=
	\begin{pmatrix}
		n_2 \\
		2n_1 + n_2
	\end{pmatrix}, \\
	\begin{pmatrix}
		\varphi_1' \\
		\varphi_2'
	\end{pmatrix}
	&=
	\begin{pmatrix}
		0 & 1 \\
		2 & 1
	\end{pmatrix}^{-T}
	\begin{pmatrix}
		\varphi_1 \\
		\varphi_2
	\end{pmatrix}
	=
	\begin{pmatrix}
		\varphi_2 - \varphi_1/2 \\
		\varphi_1/2
	\end{pmatrix},
\end{align}
\end{subequations}
and the Hamiltonian in the transformed coordinates can be written as
\begin{align}\label{eq:H_2D_trans}
	H_\text{2D}' =& \frac{e^2}{C_\text{z}}\hat{n}_1'^2 + \frac{e^2}{2C_\text{sh}+(4\alpha+1)C_\text{z}}\hat{n}_2'^2 \nonumber \\
	 -& 2 I_z \frac{\Phi_0}{2\pi} \cos(\hat{\varphi}_1')\cos(\hat{\varphi}_2'-\varphi_\text{z}/2)  \\
	 -& 2 \alpha I_z \frac{\Phi_0}{2\pi} \cos(\varphi_\text{x}/2)\sqrt{1+\tan^2(\varphi_\text{d})}\cos(2\hat{\varphi}_2' - \varphi_\text{d}). \nonumber
\end{align}

We note that in CSFQs the junction capacitance is much smaller than the shunt capacitance ($C_\text{z} \ll C_\text{sh}$), and therefore the mode corresponding to $\{ \varphi_1', n_1'\}$ has a plasma frequency that is much larger than the other mode.
Therefore, we can neglect this fast oscillating degree of freedom to reduce the number of modes in our model, i.e., we can perform a Born-Oppenheimer approximation~\cite{Born1927} which assumes the fast degree of freedom is always in its ground state.
To do this we take $C_\text{z} \rightarrow 0$ and fix $\varphi_1' = 0$, which is the phase value that minimizes the potential energy of \eqref{eq:H_2D_trans} with respect to $\varphi_1'$.
The resulting simplified Hamiltonian then becomes
\begin{subequations}
\label{eq:H_1D}
\begin{align}
	H_\text{1D} &=  K_\text{1D} + U_\text{1D} \\
	K_\text{1D} &= 	 \frac{e^2}{2C_\text{sh}}\hat{n}^2   \\
	\label{eq:U_1D}
	 U_\text{1D} &= - 2 I_z \frac{\Phi_0}{2\pi}\cos(\hat{\varphi}-\varphi_\text{z}/2)\\
	 -& 2 \alpha I_z \frac{\Phi_0}{2\pi} \cos(\varphi_\text{x}/2)\sqrt{1+\tan^2(\varphi_\text{d})}\cos(2\hat{\varphi} - \varphi_\text{d}), \notag
\end{align}
\end{subequations}
where we have dropped the subscript and prime for brevity.

\begin{table}[t]
	\begin{center}
		\begin{tabular}{|c||c|c|c|c|}
		\hline
		qubit model & $I_z$ (nA) & $C_\text{sh}$ (fF) & $C_\text{z}$ (fF) & $\alpha$\\
		\hline
		\hline
		2D CSFQ & 242 $\pm$ 3 & 62 $\pm$ 1 & 4.85 $\pm$ 0.07 & 0.423 $\pm$ 0.001 \\
		1D CSFQ & 228 $\pm$ 3 & 70 $\pm$ 1 & N/A & 0.452 $\pm$ 0.001 \\
		\hline
		\end{tabular}
	\end{center}
	\caption{Fit parameters and their $1\sigma$ standard deviation for qubit models.
	Asymmetry is fixed at $d=0.102$ for both models.
	}\label{tab:fit}
\end{table}
We call the Hamiltonian of Eq.~\eqref{eq:H_2D} the 2D model and the Hamiltonian of Eq.~\eqref{eq:H_1D} the 1D model.
We fit both of these models to our qubit spectroscopy data, which is taken by sweeping $\varphi_\text{z}$ near the degeneracy point for multiple fixed $\varphi_\text{x}$ values and measuring the resonance frequency of the microwave drive applied to the qubit through its dispersive resonator.
The spectroscopy data and the fits are shown in Supplementary Note 1, and fitted circuit parameters are presented in Table~\ref{tab:fit}.
The asymmetry is fixed at $d=0.102$ for both models, the value that is extracted via a separate measurement discussed in the results section.
To fit the 2D model of Eq.~\eqref{eq:H_2D}, we eliminate a fitting parameter by fitting only for the junction areas instead of fitting for the currents and capacitances separately, and use the design values of $3000\,\text{nA}/\mu\text{m}^2$ and $60\,\text{fF}/\mu\text{m}^2$ for the junction critical current density and capacitance density, respectively.
We note that the junction plasma frequency given our design critical current and capacitance densities is roughly 62 GHz, which is larger than the high-frequency qubit eigenstates that we used in our study.
Therefore our circuit model should remain valid for these eigenstates.

In order to find the best fit values for our multilevel circuit model, it is important to fit to spectroscopy data for the $0 \leftrightarrow 2$ transition frequency $\omega_{02}$, as well as to the $0 \leftrightarrow 1$  transition frequency $\omega_{01}$.
We also assume constant flux offsets in our model and fit for them to account for flux drifts and/or miscalibration in experiments.
The fitted values of flux offsets are smaller than a few $m\Phi_0$, which is not unexpected.
We find strong agreement between the fitted models and the experimental spectroscopy data (see Supplementary Note 1).

\section{Master Equation simulations}

To simulate the open system behavior of the qubit for linearly corrected anneal paths we use the adiabatic master equation (AME)~\cite{Albash2012}.
The system is coupled to the bath via the persistent-current operator, defined as $\hat{I}_\text{p} = - \partial U/\partial \varphi_z \times 2\pi/\Phi_0$, where $U$ is the CSFQ potential for the 2D and 1D models.
The persistent-current operator for each model is as follows:
\begin{align}\label{eq:ip}
	\hat{I}_\text{p}^\text{2D} &= I_\text{z} \cos(\hat{\varphi}_2 - \hat{\varphi}_1/2) \sin(\hat{\varphi}_1/2 - \varphi_\text{z}/2), \\
	\hat{I}_\text{p}^\text{1D} &= I_\text{z} \sin(\hat{\varphi} - \varphi_\text{z}/2).
\end{align}

The density operator of the circuit evolves according to the adiabatic master equation as
\begin{align}
	\dot{\rho} =& -i[H + H_{LS}, \rho] \nonumber \\
	&+ \sum_\omega \gamma(\omega) \left[ L_\omega \rho L_\omega^\dagger -\frac{1}{2} \{ L_\omega^\dagger L_\omega, \rho \}\right].
\end{align}
where
\begin{equation}
	\gamma(\omega) = \eta g^2 \frac{2\pi\omega e^{-|\omega|/\omega_c}}{1-e^{-\beta\omega}}
\end{equation}
is the Ohmic bath spectral function, with a high frequency cut-off at $\omega_c/2\pi = 15\,\text{GHz}$, and is in thermal equilibrium at $T=1/k_BT=10~\text{mK}$.
Conforming to the notations in Ref.~\cite{Albash2012}, $\eta g^2=3\times 10^{-6}$ is the system-bath coupling strength, where $\eta g^2/\hbar$ has units of $1/\text{energy}^2$.
The Lindblad operators are calculated as
\begin{equation}
	L_\omega = \sum_{\varepsilon_b-\varepsilon_a=\omega} \bra{\varepsilon_a} I_\text{p} \ket{\varepsilon_b}\ket{\varepsilon_a}\bra{\varepsilon_b}
	= L_{-\omega}^\dagger,
\end{equation}
where $\varepsilon_k$ and $\ket{\varepsilon_k}$ are eigenvalues and eigenvectors of the Hamiltonian respectively.
$H_{LS}$ denotes the Lamb shift, which is calculated as
\begin{equation}
	H_{LS} = \sum_\omega L_\omega^\dagger L_\omega S(\omega),
\end{equation}
with
\begin{equation}
	S(\omega) = \int_{-\infty}^{\infty} \frac{d\omega'}{2\pi} \gamma(\omega)\mathcal{P}\left(\frac{1}{\omega-\omega'}\right),
\end{equation}
where $\mathcal{P}$ denotes the Cauchy principal value.

The AME formalism breaks down if one tries to replace the Ohmic bath spectral function with a $1/f$ spectrum~\cite{Albash2012}.
To handle this case other tools are needed, such as recent work on open-system evolution equations that can capture the effects of both fast and slow noise~\cite{Smirnov2018,Mozgunov2019}.
We have evidence (work in progress) that the polaron-transformed Redfield equation~\cite{Xu2016} with hybrid (slow and fast) environments yields linewidth-broadened features compared to the AME.

We note that in order to keep the computations for the multi-level circuit manageable, at each time step of the ODE solver we rotate the density matrix into the instantaneous eigenbasis of the Hamiltonian that is truncated (e.g., truncated at 10 eigenlevels), calculate all the above terms for AME, and then rotate it back into its initial basis.

\section{Data Availability}
The data supporting the findings of this study are available within the paper.
The data are available from the authors upon reasonable request and with the permission of our US Government sponsors.
\section{Code Availability}
The codes that support the findings of this study are available from the authors upon reasonable request and with the permission of our US Government sponsors.

\begin{acknowledgments}
\section{Acknowledgments}
We are grateful to David G. Ferguson for insightful discussions and to all members of the Quantum Enhanced Optimization (QEO) team for their collaboration, especially MIT Lincoln Laboratory.
The research is based upon work supported by the Office of the Director of National Intelligence (ODNI), Intelligence Advanced Research Projects Activity (IARPA) and the Defense Advanced Research Projects Agency (DARPA), via the U.S. Army Research Office contract W911NF-17-C-0050. 
The views and conclusions contained herein are those of the authors and should not be interpreted as necessarily representing the official policies or endorsements, either expressed or implied, of the ODNI, IARPA, DARPA, or the U.S. Government.
The U.S. Government is authorized to reproduce and distribute reprints for Governmental purposes notwithstanding any copyright annotation thereon.
\end{acknowledgments}

\section{Competing Interests}
The authors declare no competing interests.
\section{Author Contributions}
M.K. performed the theoretical analysis and simulations, proposed the asymmetry correction procedure, and explained the level crossing data.
J.G. performed all the experiments and analyzed the data.
J.G., J.B., S.D, and S.N. built the experimental setup.
H.C. wrote the open system simulation codes.
M.K. and J.G. wrote the initial version of the manuscript.
D.L. revised it, and all other authors helped with final revisions.
K.Z. guided the experimental team and D.L. guided the entire project.
M.K. and J.G. contributed equally to this work.

\newpage

\clearpage 
\onecolumngrid 
\vspace{\columnsep}
\begin{center}
\textbf{\large Supplementary Information for} \\
\vspace{0.2cm}
 \textbf{\large Anneal-path correction in flux qubits}
\end{center}
\vspace{\columnsep}
\twocolumngrid

\setcounter{equation}{0}
\setcounter{figure}{0}
\setcounter{table}{0}
\setcounter{page}{1}

\renewcommand{\theequation}{S\arabic{equation}}
\renewcommand{\thefigure}{S\arabic{figure}}
\renewcommand{\thetable}{S\arabic{table}}
\renewcommand{\bibnumfmt}[1]{[S#1]}
\renewcommand{\citenumfont}[1]{S#1}

\renewcommand{\theHtable}{Supplement.\thetable}
\renewcommand{\theHfigure}{Supplement.\thefigure}
\renewcommand{\theHequation}{Supplement.\theequation}

\section*{Supplementary Note 1: Spectroscopy data and fit}
\begin{figure}[t]
	\begin{center}
		\includegraphics[width=\columnwidth]{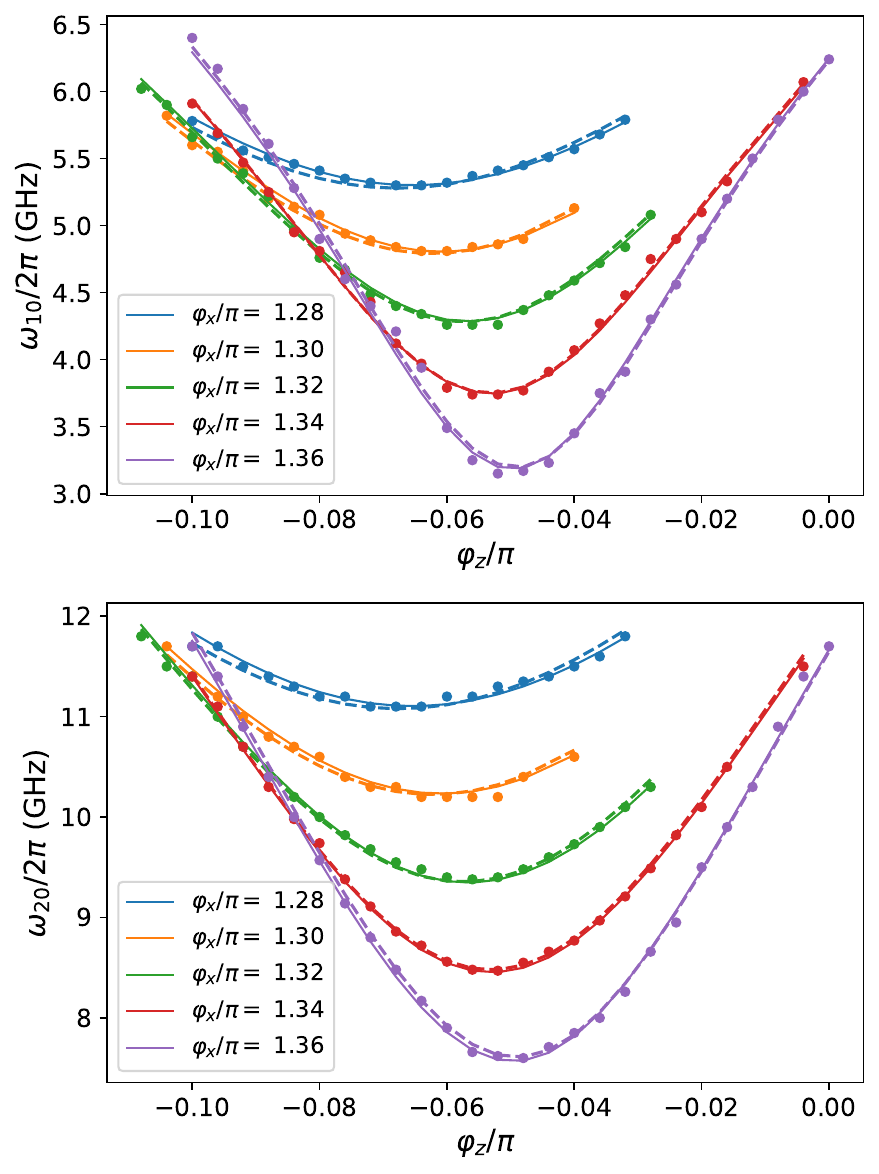}
    \end{center}
    \caption{
	Fit and experimental spectroscopy data for the $\omega_{10}$ (top panel) and $\omega_{20}$ (bottom panel) transition frequencies.
	Filled circles are qubit resonance frequencies extracted from experimental spectroscopy data.
	Solid lines are the fit to the 2D qubit model, and dashed lines are the fit to the 1D qubit model.
	Each color matched band corresponds to a spectroscopic measurement where $\varphi_\text{x}$ was kept fixed and $\varphi_\text{z}$ was swept near the degeneracy point.
    }\label{fig:spec_fit}
\end{figure}

Fig.~\ref{fig:spec_fit} shows our qubit spectroscopy data (filled circles), and strong agreement between the fitted circuit models and the data.

\begin{figure}[t]
	\begin{center}
		\includegraphics[width=\columnwidth]{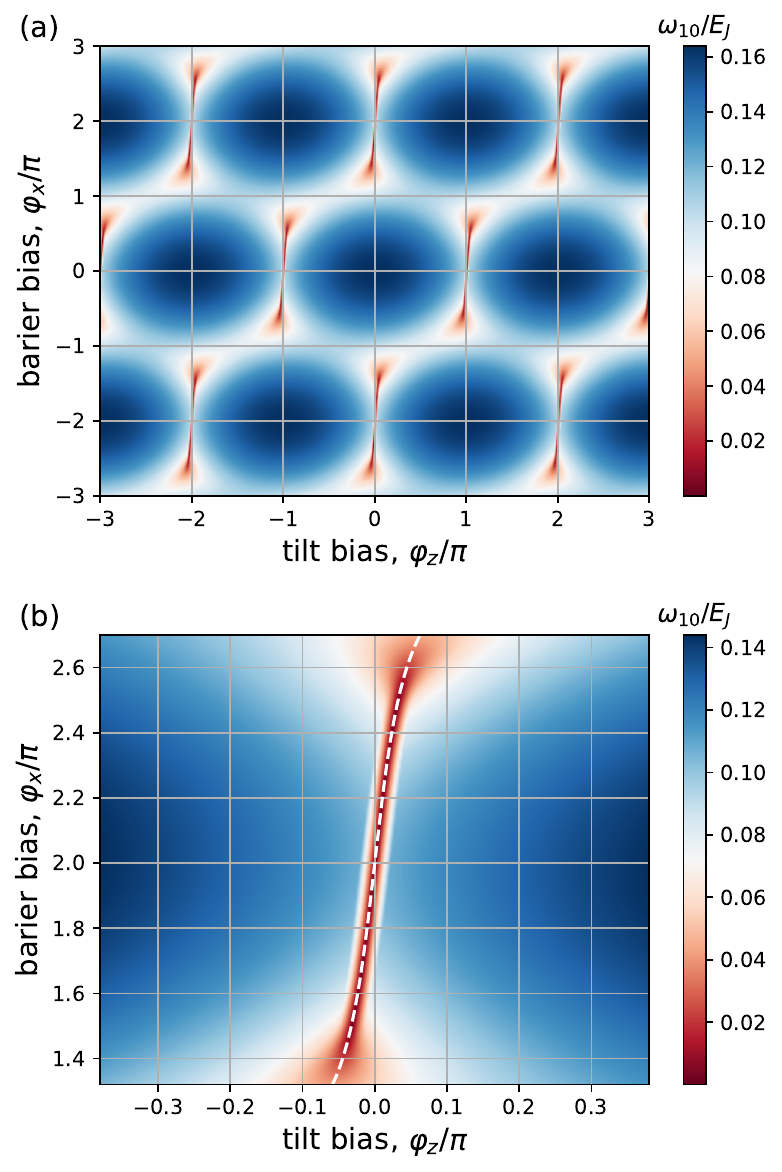}
    \end{center}
    \caption{
	Gap of the qubit as a function of control fluxes, calculated by diagonalizing the circuit Hamiltonian using the fitted parameters of the circuit, presented in Table~II of the main text.
	For our parameters, $E_\text{J}/2\pi \approx 100~\text{GHz}$
	(a) Periodic structure of the gap, showing multiple ``cells''.
	(b) Zoomed-in gap, with bias ranges that correspond to Fig.~3 of main text that was used for extraction of the asymmetry parameter.
	White dashed line shows the location of the minimum gap, which occurs at $\varphi_\text{z} = \varphi_\text{d}$ for $\pi \leq \varphi_\text{x} \leq 3\pi$.
    }\label{fig:gap}
\end{figure}

With all the circuit parameters extracted via fit to spectroscopy data, we can numerically calculate the $0\leftrightarrow1$ transition frequency of the circuit as a function of control biases.
The result is shown in Fig.~\ref{fig:gap}.
The qubit gap is a periodic function of the control fluxes, and annealing paths could be chosen from any of the periodic ``cells'' [Fig.~\ref{fig:gap}(a)].
The asymmetry extraction procedure discussed in the main text uses one of these cells, which is shown in Fig.~\ref{fig:gap}(b) for the same flux ranges as in Fig.~3 of the main text.
It can be seen that the minimum gap occurs at $\varphi_\text{z} = \varphi_\text{d}$ for $\pi \leq \varphi_\text{x} \leq 3\pi$, indicated by the white dashed line in Fig.~\ref{fig:gap}(b).
Note that at $\varphi_\text{x}=0$, the degeneracy occurs at $\varphi_\text{z}=\pi$ (i.e., half flux-quantum).

\section*{Supplementary Note 2: Persistent-current readout}

The persistent-current readout uses a quantum flux parametron (QFP), which is positioned between the qubit and an rf-SQUID resonator and inductively coupled to both (Fig.~1 of main text).
The QFP, which is a larger flux-qubit-like device operated in a classical regime, amplifies the persistent-current signal and isolates the CSFQ from the resonator.
This reduces the Purcell effect and increases $T_1$~\cite{Strand2019, Grover2020S}.
At the end of each anneal, the circulating current in the qubit creates an effective tilt bias on the QFP that changes the direction of its circulating current, which in turn shifts the rf-SQUID resonator frequency that can be measured to infer the direction of the circulating currents.

The persistent-current readout has an effective positive operator valued measure (POVM) for calculating the probability of measuring the right circulating current, which can be written as
\begin{equation}
	\hat{M}_r = \sum_\lambda\, f\! \left( \frac{I_\lambda}{\Delta I} \right) \ket{\lambda}\bra{\lambda},
\end{equation}
where $I_\lambda$ and $\ket{\lambda}$ are the eigenvalues and eigenvectors of the persistent-current operator respectively, $f(x) = [\tanh(x)+1]/2$ is a filter function, and $\Delta I$ is the sensitivity of the persistent-current readout device, which in our QFP-based system is $\Delta I = 10\,\text{nA}$.
The probability of measuring the right circulating current is then $P_r = \text{Tr}(\rho \hat{M}_r)$, where $\rho$ is the qubit density matrix.

\section*{Supplementary Note 3: Mapping Circuit to Ising spin}

In order to map the multi-level circuit Hamiltonian into an Ising spin model with only two levels, we keep the two lowest eigenenergies of the CSFQ, as these are the two levels we use for representing a qubit.
Furthermore, because we perform a persistent-current measurement at the end of each anneal, we would like the computational basis to be the eigenstates of the persistent-current operator.
Therefore, we first write the persistent-current operator in the low-energy subspace as
\begin{equation}
	I_\text{p}^\text{low} = 
	\begin{pmatrix}
		\langle g| \hat{I}_\text{p} |g\rangle & \langle g| \hat{I}_\text{p} |e\rangle \\
    		\langle e| \hat{I}_\text{p} |g\rangle & \langle e| \hat{I}_\text{p} |e\rangle
\end{pmatrix},
\end{equation}
where $\{ \ket{g}, \ket{e}  \}$ are the ground and exited eigenstates of the \emph{circuit} Hamiltonian with eigenenergies $\{ E_g, E_e \}$ respectively, and $\hat{I}_p$ is the persistent-current operator.

Note that for flux qubits where we associate the qubit states to circulating currents flowing in opposing directions, we require the eigenvalues of $I_\text{p}^\text{low}$ to have opposite signs.
If we tilt the qubit potential beyond a certain point, then the first two eigenstates of the circuit will both be localized in the same well and the eigenvalues of $I_\text{p}^\text{low}$ will have the same sign.
This puts an upper bound on $|\varphi_\text{z}|$, which is illustrated in Fig.~2 of the main text.

Now let $U$ be the unitary basis transformation that diagonalizes $I_\text{p}^\text{low}$, or in other words, transforms the energy basis into the computational (persistent-current) basis.
$U$ is formed from the eigenstates of $I_\text{p}^\text{low}$ as its columns.
The computational basis $\{ \ket{0}, \ket{1}  \}$ is then
\begin{align}
	\begin{pmatrix}
		|0\rangle  \\
		|1\rangle\\
	\end{pmatrix}
 &= U^\dagger 
	\begin{pmatrix}
		|g\rangle  \\
		|e\rangle\\
	\end{pmatrix} ,
\end{align}
and the effective Hamiltonian in the computational basis is 
\begin{equation}
	H_\mathrm{eff} = U^\dagger
	\begin{pmatrix}
		E_g & 0 \\
		0 & E_e
	\end{pmatrix}
	U.
\end{equation}
We extract the Ising coefficients by rewriting the effective Hamiltonian as
\begin{equation}
	H_\mathrm{eff} = \alpha_x\sigma_x + \alpha_y\sigma_y + \alpha_z\sigma_z + \alpha_I I .
\end{equation}

For simplicity, the following constraints are usually imposed on the effective Hamiltonian by applying an additional unitary transformation to the computational basis:
\begin{enumerate}
	\item $\alpha_y$ is set to zero.
	\item $\alpha_x$ is always positive.
\end{enumerate}
After imposing the above constraints, we can write the effective Hamiltonian as a standard transverse field Ising Hamiltonian of the form
\begin{equation}
	H_\mathrm{eff} = A \sigma_x + B \sigma_z.
\end{equation}
This procedure leads to Fig.~2 in the main text.

\begin{figure}[t]
	\begin{center}
		\includegraphics[width=0.9\columnwidth]{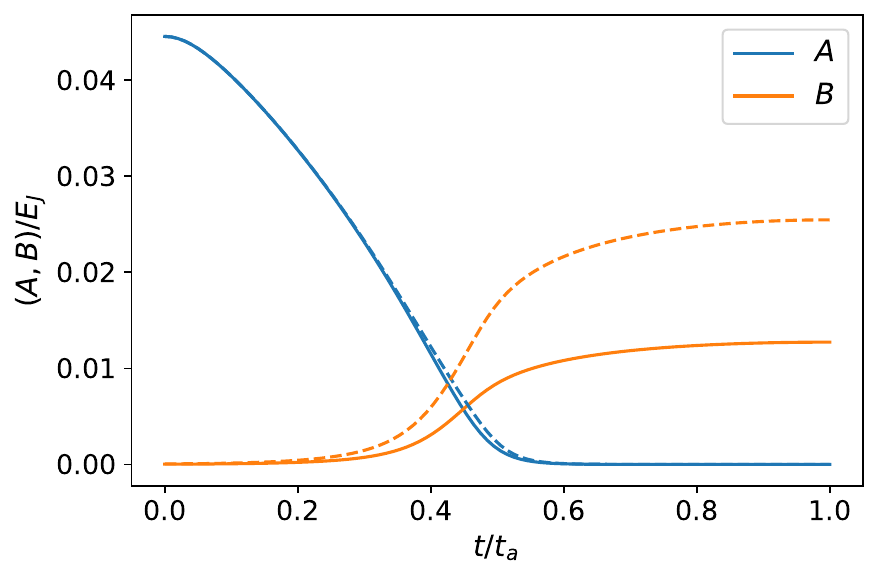}
    \end{center}
    \caption{
	Transverse field Ising Hamiltonian coefficients for two of the asymmetry-{corrected} anneal paths, \textit{vs} normalized anneal time $t/t_a$.
	Solid lines correspond to the anneal path with $\varphi_\text{z}(0)/\pi = 0.005$, dashed lines correspond to the path with $\varphi_\text{z}(0)/\pi = 0.01$.
	For our system $E_\text{J}/2\pi \approx 100~\text{GHz}$.
    }\label{fig:AB_s}
\end{figure}

As an example, and to make the connection between the s-curve measurements and the qubit picture, we calculate the $A$ and $B$ coefficients for two of the asymmetry-{corrected} anneal paths that were used for the measurements in Fig.~4 of the main text.
The result is shown in Fig.~\ref{fig:AB_s}, where the solid lines correspond to the anneal path with $\varphi_\text{z}(0)/\pi = 0.005$, and the dashed lines correspond to the path with $\varphi_\text{z}(0)/\pi = 0.01$.

\section*{Supplementary Note 4: Experimental setup}
\label{sec:setup}
\begin{figure*}[t]
	\begin{center}
		\includegraphics[width=0.9\textwidth]{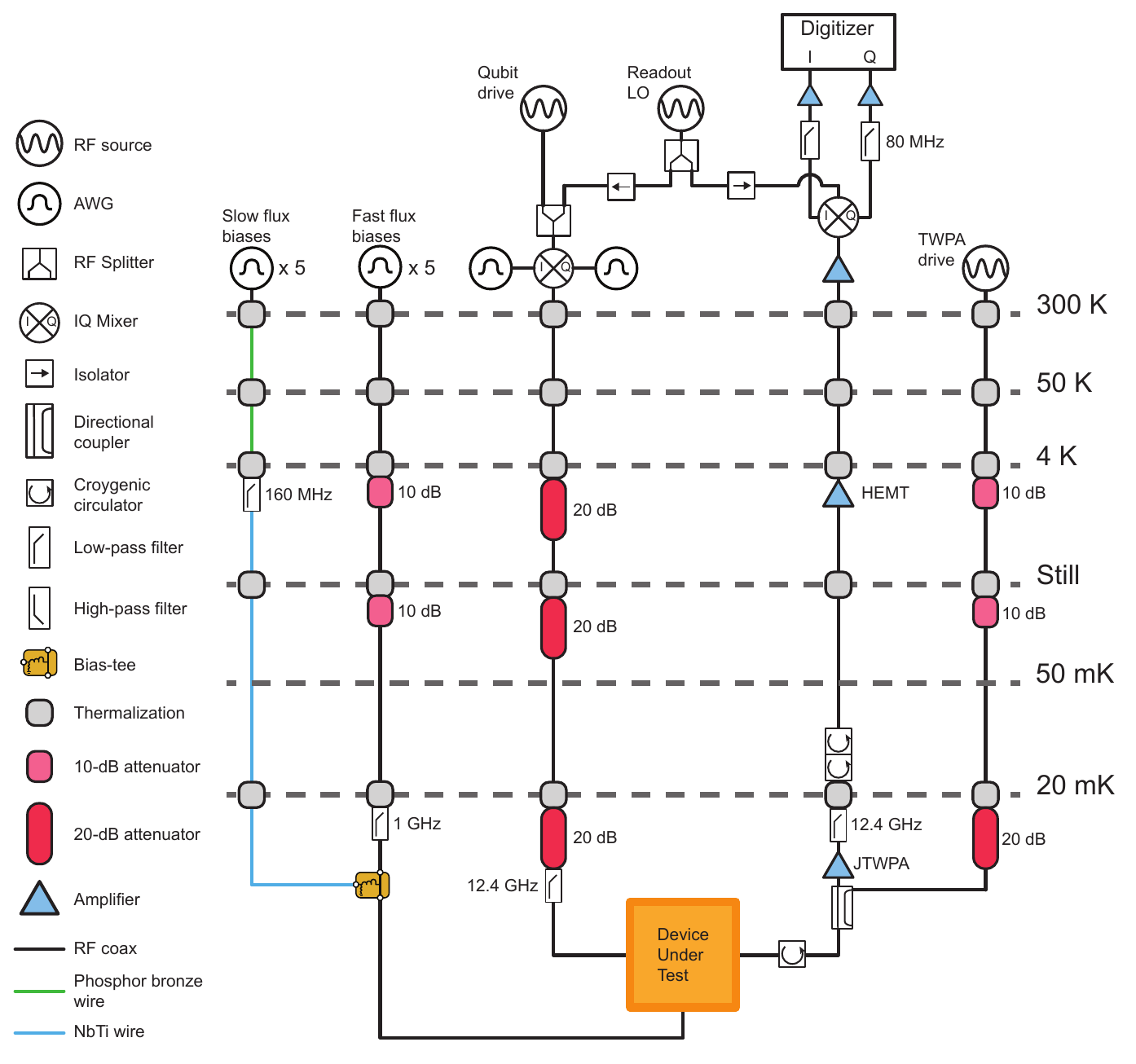}
	\end{center}
	\caption{
	A schematic diagram of the room-temperature measurement setup and dilution refrigerator wiring.
	}
	\label{fig:fridge}
\end{figure*}

Figure~\ref{fig:fridge} is a schematic representation of the measurement setup from room temperature down to the mixing chamber.
The experiments are performed in a Leiden Cryogenics dilution refrigerator, with a base temperature in the range of $15-25$ mK.
Slow flux biases are provided by independent arbitrary waveform generator (AWG) channels and reach the device via phosphor-bronze ribbon cables from 300 K to 4 K, followed by NbTi cables from 4 K to the mixing chamber.
Fast control is provided by different, independent AWG channels that utilize their full 1 GS/s time resolution.
These fast biases are sent down coax lines, and they are combined with the slow biases at the mixing chamber via cryogenic bias-tees with an added 1-GHz low-pass filter.
Output signals are first amplified by a Josephson traveling-wave parametric amplifier (JTWPA)~\cite{Macklin2015} at the mixing chamber, followed by a high-electron-mobility transistor (HEMT) amplifier anchored at $4$ K.

For readout we use a split-heterodyne configuration with image rejection to downconvert signals into the intermediate frequency (IF) band, typically 50 MHz.
A field programmable gate array (FPGA) digitizer performs analog-to-digital conversion for signal processing and analysis.
A simple boxcar windowing function is applied for IF demodulation~\cite{Krantz2019}, which is done either directly on the FPGA or in software after the full signal traces are transferred off the card.

Details of the device layout and fabrication can be found in the Appendix of Ref.~\cite{Grover2020S}.

\section*{Supplementary Note 5: Lifetime and Coherence}
\label{sec:lifetime}
\begin{figure}[t]
	\begin{center}
		\includegraphics[width=\columnwidth]{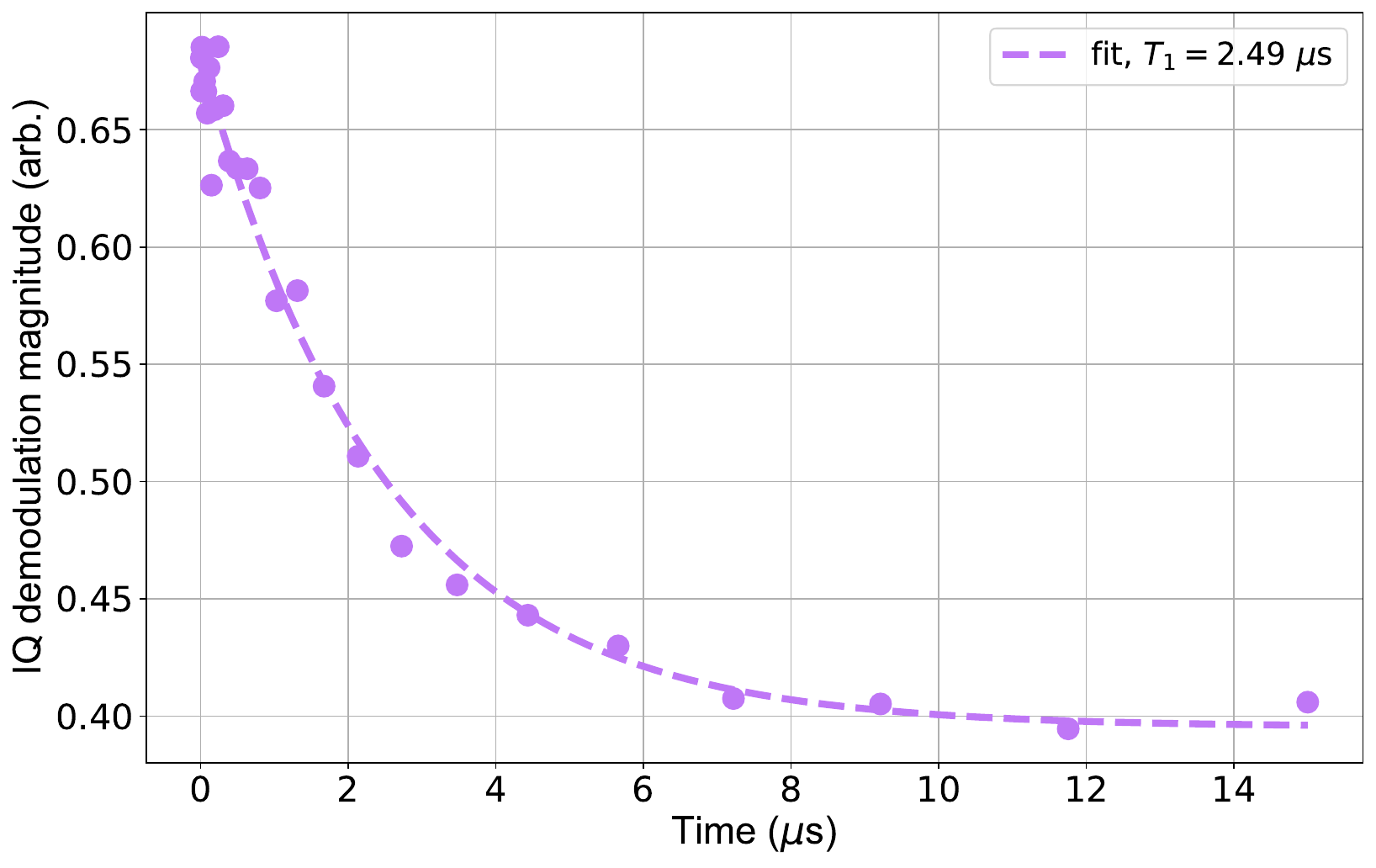}
    \end{center}
    \caption{
    One realization of an energy relaxation measurement of the qubit at a frequency of 5.9 GHz.
    The plot uses IQ demodulation amplitude on the y-axis, which is proportional to excited-state population.
    The fit (dashed purple line) is a simple exponential decay model, and we obtain a lifetime of $2.49$ $\mu$s for this dataset.
    }\label{fig:t1}
\end{figure}

Figure~\ref{fig:t1} presents a characteristic lifetime measurement of the qubit at a transition frequency of 5.9 GHz.
Repeated measurements yield an average value of $T_1 = 1.93 \pm 0.19$ $\mu$s ($1\sigma$ error).
We have confirmed that qubit lifetime remains roughly constant over a range of qubit frequencies of about 1 GHz around this measurement, but it is expected that lifetime will decrease as the gap closes~\cite{Quintana2017s}.
Using Ramsey interferometry, we measured a dephasing time of $T^{*}_{2} = 130$ ns at a qubit frequency of 4.2 GHz in a similar device (four-junction CSFQ with the same design $I_\text{p}$) from the same fabrication run~\cite{Novikov2018S}.

We note that $T_1$ is affected by a combination of slow flux noise, fast charge noise, and Purcell decay through the readout resonator, and the qubit dephasing time (and therefore $T_2$) can be adversely affected by the shot noise of residual photons in the dispersive resonator.
In general, as the qubit gap becomes small, the slow flux noise becomes dominant and degrades the energy relaxation time with a scaling of $1/I_\text{p}^2$ and the coherence time with a Gaussian decay scaling of $1/I_\text{p}$.
At large qubit frequencies the fast charge noise takes over and degrades the lifetime, although it is sometimes challenging to distinguish it from fast flux noise \cite{Yan2016S}.
Additionally, as the detuning between the qubit and the readout resonator becomes small, the qubit lifetime degrades due to an increase in the Purcell decay \cite{Blais2004}.

The interplay between all these effects on CSFQs was extensively studied in Ref.~\cite{Yan2016S}, where the coherence times and their main contributing factors varied depending on different qubit operating regimes and system parameters.
A similar systematic study of coherence in the present system was beyond the scope of the our work.
Nevertheless, due to the similarity in CSFQ design and parameters and fabrication, we expect similar results to those of Ref.~\cite{Yan2016S}.
\section*{Supplementary Note 6: The effect of idling post anneal}

\begin{figure}[t]
	\begin{center}
		\includegraphics[width=\columnwidth]{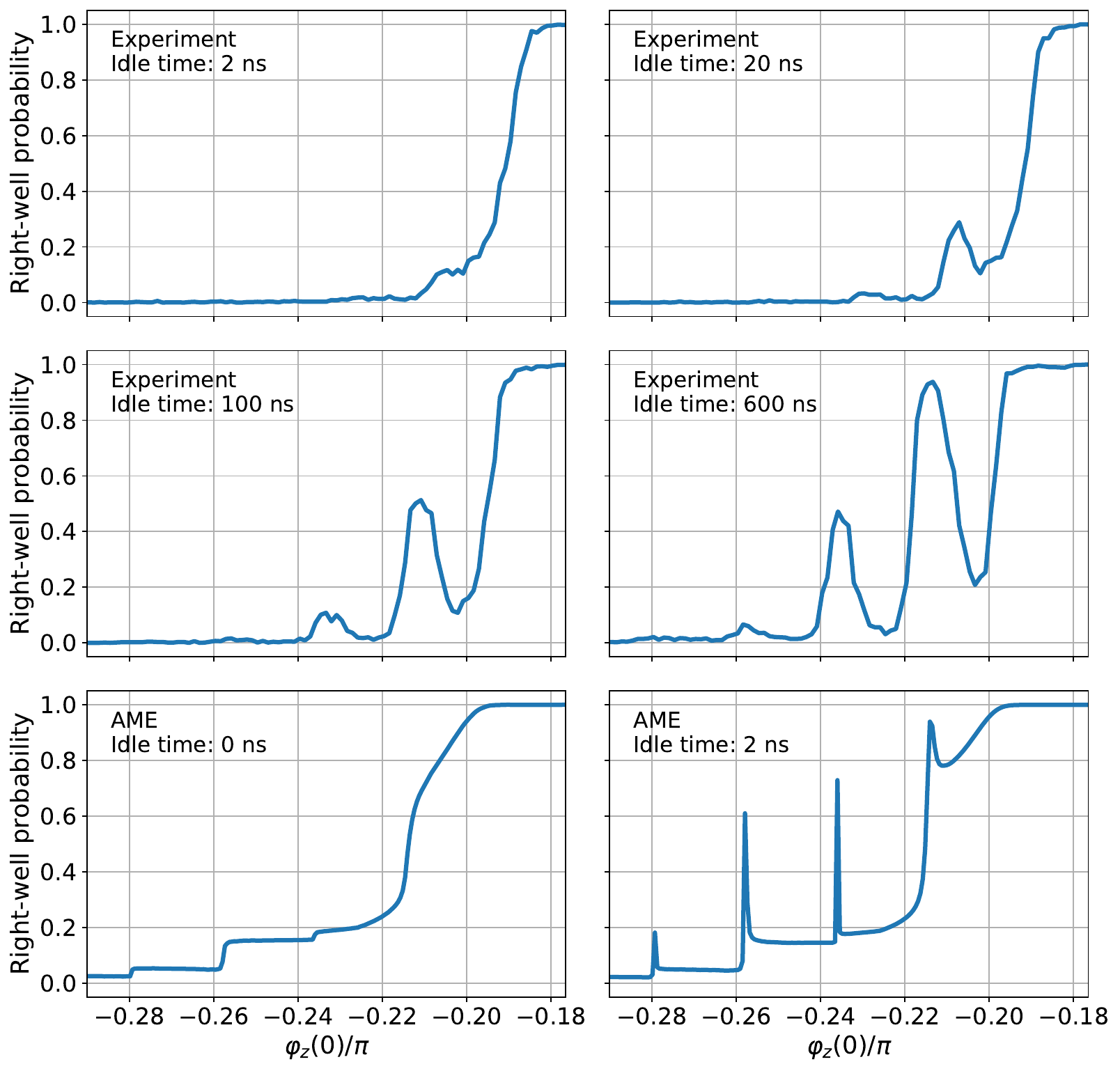}
    \end{center}
    \caption{
    The effect of post-anneal idle time on s-curve characteristics.
    The anneal is similar to the one described in Fig.~5 of the main text: the anneal occurs in $60$~ns, with an amplitude of $0.33\pi$ applied to the tilt bias.
    From left-to-right and top-to-bottom, the first four panels show the experimental results when the idle time before readout is increased from $2$~ns to $600$~ns. The last two panels show the effect of idle time in AME simulations.
    The fast rise of the peaks in AME simulations is expected (see the text).
    }\label{fig:idle}
\end{figure}

As illustrated in Fig.~5 of the main text, peaks in the s-curve appear when the anneal path traverses level crossings, leading to diabatic population transfer.
The AME simulations reproduce these peaks only when an idle time is added between the end of the qubit anneal and readout.
Without any delay, the theory predicts that instead the s-curve will exhibit plateaus.
This delay allows for relaxation to occur, redistributing population between levels in either well.
However, note that the AME with an Ohmic bath produces large transition rates at small gaps \cite{Albash2015}, which means the peaks will rise faster than they do in the experiment.
Nevertheless, the AME can qualitatively predict the effect of an idle time after the anneals, as seen in Fig.~\ref{fig:idle} and Fig.~5(a) of the main text.

We confirm this effect experimentally by varying the idle time after the anneals, as shown in Fig.~\ref{fig:idle}.
The anneal is depicted in Fig.~3 of the main text, where the anneal traverses a ``tilted'' path in flux space due to a large amplitude applied to the tilt bias.
As the idle time increases from $2$~ns to $600$~ns, plateau-like features in the s-curve become peaks, as expected from the theory.

We performed similar delay studies with asymmetry-corrected anneal paths, and neither the plateaus nor the peaks appeared (not shown).
This suggests that fewer excitations into higher-energy states occurred.

\section*{Supplementary Note 7: Experimental pulses}
In this section we describe in more detail the actual pulses used to perform the correction, as well as how different pulse parameters affect the s-curve width.

Note that for experimental parameters, we use real flux values in units of $\Phi_0$, and recall the relation to phase: $\varphi_{\text{x},\text{z}}=2\pi\Phi_{\text{x},\text{z}}/\Phi_0$.
As implied by Eq.~(2) in the main text, we parametrize the $z$-flux in terms of the $x$-flux.
We first decide on a functional form and duration for $\Phi_\text{x} (t)$, and then Eq.~(2) is used to determine $\Phi_\text{z} (t)$ from those values of $\Phi_\text{x}$.
We found that a gaussian pulse shape for the $x$-flux produced better results than a linear ramp, likely due to reduced pulse distortion in the RF-coax lines.
Future studies could explore using more sophisticated techniques, such as DRAG~\cite{Motzoi2009} or optimal control~\cite{Werninghaus2020}, to find more performant pulses.

\begin{figure}[t]
	\begin{center}
		\includegraphics[width=\columnwidth]{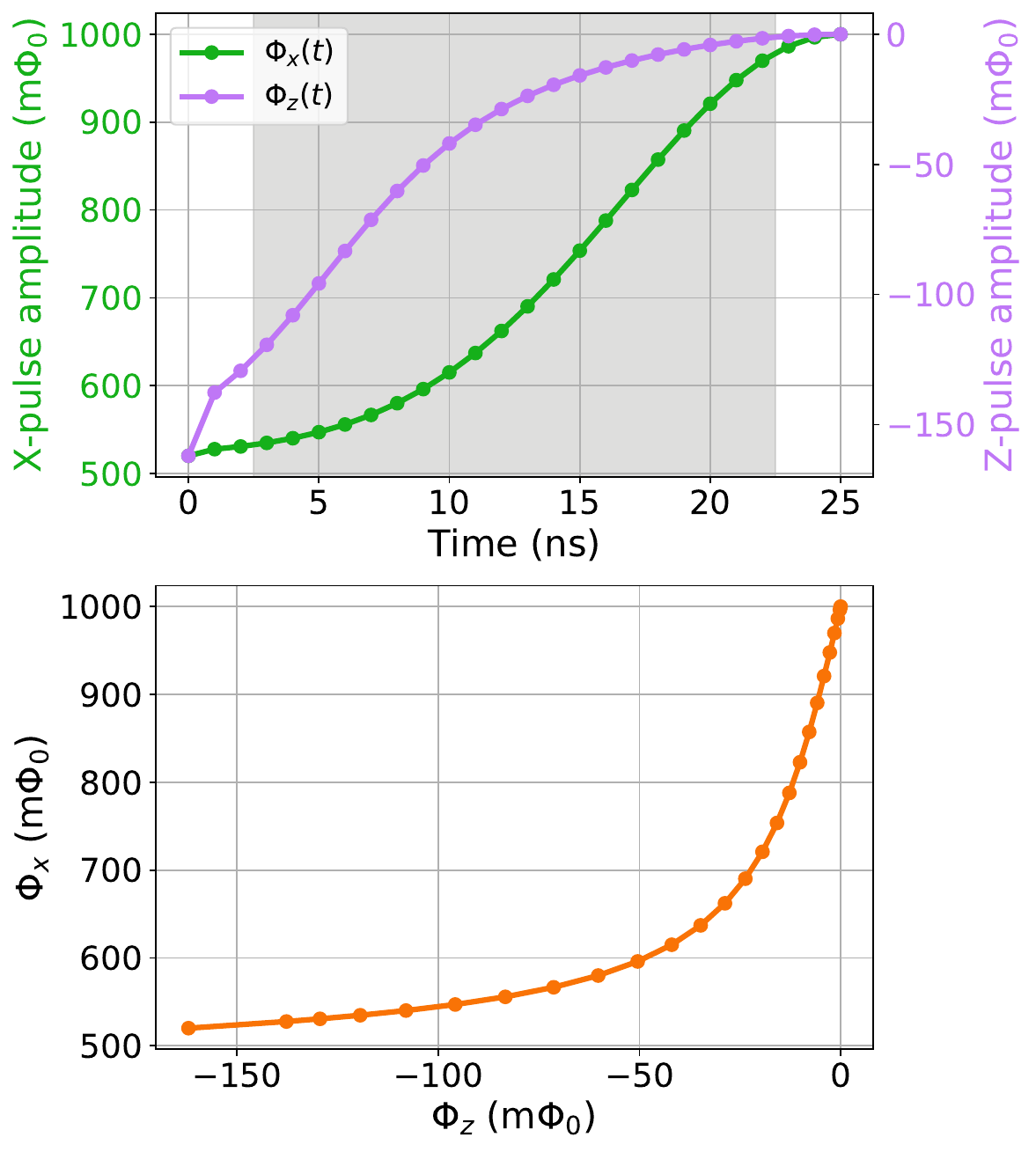}
    \end{center}
    \caption{
    Experimental pulses used to correct for junction asymmetry of magnitude $d=0.102$.
    Top: $x$ [$\Phi_\text{x} (t)$, green circles] and $z$ [$\Phi_\text{z} (t)$, purple circles] flux pulses at the outputs of the AWGs as a function of time.
    The hitch after the first point is due to truncation of the gaussian pulse, and can be smoothed in software.
    The gray shaded area indicates the $20$~ns rise time.
    Bottom: The corrected pulse visualized in flux space, where each orange circle corresponds to one time step from the top plot.
    }\label{fig:pulses}
\end{figure}

\begin{figure}[t]
\vspace{1cm}
	\begin{center}
		\includegraphics[width=\columnwidth]{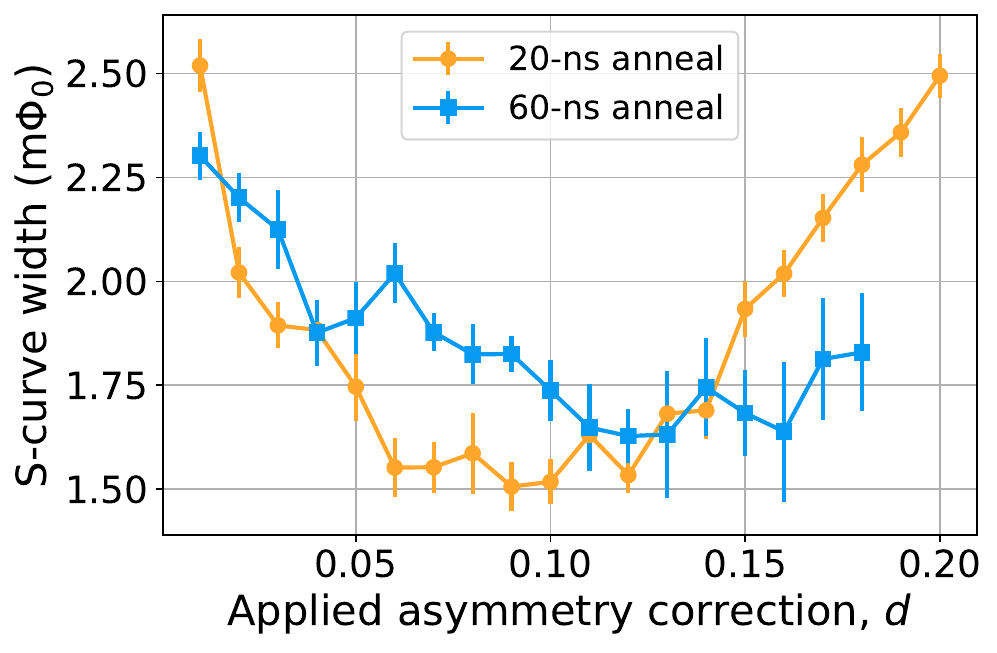}
    \end{center}
    \caption{
    S-curve width versus applied correction parameter for two different anneal times of $20$~ns (orange circles) and $60$~ns (blue squares).
    }\label{fig:vary_d}
\end{figure}

Figure~\ref{fig:pulses} illustrates the pulses used to correct for an asymmetry of $d=0.102$.
The top plot shows each independent control line as a function of time.
The shaded gray area indicates the $5-95\%$ rise time of $20$~ns, which is the parameter used when defining the ``anneal time'' for a particular experiment.
The bottom plot then combines these two pulse to illustrate the actual annealing path traversed by the qubit.
The displayed pulse shapes are as-measured at the ouputs of the $1$ GS/s arbitary waveform generators (AWGs).
In this study, no pulse distortion correction was applied~\cite{Gustavsson2013,Rol2020}, but it will be part of future experimental improvements.

We also studied the effect of changing the value of $d$ when applying the correction.
Fig.~\ref{fig:vary_d} highlights the reduction in s-curve width when the value of $d$ is near the value of $\approx 0.102$ measured during calibration (and verified in simulation).
The difference in trends between the two anneal times suggests that there is potentially more room for optimization by exploring the entire space of annealing time in addition to the path trajectory.

We also note that the data in Fig.~\ref{fig:vary_d} were taken when the fridge base temperature was $20$\% higher ($24$~mK) than that in the main text ($20$~mK).
This accounts for the overall increase in measured s-curve widths.

\FloatBarrier

\end{document}